# Domain Wall Enabled Steep Slope Switching in $MoS_2$ Transistors Towards Hysteresis-Free Operation

Jingfeng Song[1], Yubo Qi[2], Zhiyong Xiao[1], Kun Wang[1], Dawei Li[1], Seung-Hyun Kim[3], Angus I. Kingon[3], Andrew M. Rappe[2] and Xia Hong[1*]

[1] *Department of Physics and Astronomy, University of Nebraska-Lincoln, Lincoln, NE 68588, USA*

[2] *Department of Chemistry, University of Pennsylvania, Philadelphia, PA 19104-6323, USA*

[3] *School of Engineering, Brown University, Providence, RI 02912, USA*

[*] email: xia.hong@unl.edu

## Abstract

The device concept of ferroelectric-based negative capacitance (NC) transistors offers a promising route for achieving energy-efficient logic applications that can outperform the conventional semiconductor technology, while viable operation mechanisms remain a central topic of debate. In this work, we report steep slope switching in $MoS_2$ transistors back-gated by single-layer polycrystalline $PbZr_{0.35}Ti_{0.65}O_3$. The devices exhibit current on/off ratios up to $8\times10^6$ within an ultra-low gate voltage window of $V_g = \pm0.5$ Vand subthreshold swing (SS) as low as 9.7 mV decade$^{-1}$ at room temperature, transcending the 60 mV decade$^{-1}$ Boltzmann limit without involving additional dielectric layers. Theoretical modeling reveals the dominant role of the metastable polar states within domain walls in enabling the NC mode, which is corroborated by the relation between SS and domain wall density. Our findings shed light on a hysteresis-free mechanism for NC operation, providing a simple yet effective material strategy for developing low-power 2D nanoelectronics.

## Introduction

While the ever-growing thermal power becomes a central challenge faced by information technology in the post-Moore's law era[1], ferroelectric-gated field effect transistors (FeFETs) operating in the negative capacitance (NC) mode provides a promising route for developing energy-efficient logic applications that can transcend the classic thermal limit[2, 3]. For conventional transistors, the subthreshold swing (SS), defined as the gate voltage ($V_g$) required to change the channel source-drain current $I_d$ by one order of magnitude (decade, dec), is determined by Boltzmann statistics:

$$\mathrm{SS} \equiv \frac{\partial V_\mathrm{g}}{\partial(\log_{10} I_\mathrm{d})} = \frac{\partial V_\mathrm{g}}{\partial \psi_\mathrm{s}} \cdot \frac{\partial \psi_\mathrm{s}}{\partial(\log_{10} I_\mathrm{d})} = \left(1 + \frac{C_\mathrm{ch}}{C_\mathrm{g}}\right)\frac{k_\mathrm{B}T}{q}\ln 10\ , \qquad (1)$$

which imposes a fundamental limit of SS $\approx 60$ mV dec$^{-1}$ at 300 K[2]. Here $\psi_s$ is the surface potential of the channel, $C_{ch}$ is the channel capacitance, and $C_g$ is the gate capacitance. It has been proposed that by replacing the gate dielectric with a ferroelectric layer coupled with proper capacitance matching, it is possible to stabilize the device in the regime with an effectively negative $C_g$, which in turn reduces SS below the Boltzmann limit (Equation (1)), known as steep slope switching[2]. The key to accessing the intrinsic NC regime of ferroelectrics relies on the instability of the spontaneous polarization[2], which has been identified experimentally either in single-layer ferroelectric capacitors in transient measurements during polarization switching[4, 5], or in ferroelectric/dielectric stacks exploiting the dielectric layer to stabilize the quasi-static NC mode[6-15]. Since polarization switching is a first-order physical process, a hysteresis loop in the transfer curve $I_d(V_g)$ is inevitable, which means that the on and off switching must be operated at different voltages. Such hysteresis window is not desired as it effectively increases the turn-on voltage span, lowers operation speed, and compromises the reliability of the device performance. Alternative scenarios proposed to harness the NC effect include charge trapping[16] and polarization rotation effects[17]. While the underlying mechanism for the NC-FETs remains a central topic of debate, the technological implementation of this device concept calls for device switching in a hysteresis-free fashion[3].

Since the initial proposal of the NC-FET, a wide variety of material systems have been investigated theoretically or experimentally as channel materials for NC-FET. Compared with conventional semiconductors[6, 7], the two-dimensional (2D) layered transition metal dichalcogenides (TMDCs) such as $MoS_2$ and $MoSe_2$[8-14, 18] offer an intrinsic advantage in terms of size scaling[19]. Mono- to few-layer $MoS_2$ is a semiconductor with band gap of 1.2−1.8 eV, and has been widely investigated for building high-performance logic applications[19], where high current on/off ratio[20, 21], high mobility[22], and high breakdown field[23] have been demonstrated using conventional dielectric gates, such as $SiO_2$ and $HfO_2$[20-23]. Interfacing TMDCs with ferroelectric oxides[9-13, 24-26] and polymers[8, 14, 18, 27-29] further introduces a plethora of functionalities into the 2D channel, including nonvolatile memories, programmable junctions, and steep slope transistors[3, 30].

In this work, we report steep slope switching in few-layer (FL) and bilayer (2L) $MoS_2$ transistors back-gated by single-layer polycrystalline $PbZr_{0.35}Ti_{0.65}O_3$ (PZT) films. These devices exhibit current ratios up to $8\times10^6$ within an ultra-low gate voltage window of $V_g = \pm 0.5$ V, SS as low as 9.7 mV dec$^{-1}$ at room temperature, and hysteresis-free switching at $I_d > 10^{-12}$ A μm$^{-1}$. Unlike the extensively investigated device structure with a ferroelectric/dielectric stack gate, no dielectric layer is employed to stabilize the NC mode of the ferroelectric layer. Instead, our theoretical modeling reveals that the steep slope switching originates from the metastable polar state within the domain walls (DWs) in the polycrystalline PZT gate, where a sudden boost of surface potential can be induced at an electric field well below the ferroelectric coercive field. Compared with conventional NC mechanisms that involve polarization switching, the operation based on polarizing the DW is intrinsically low power and should occur at high speed. Our study thus provides critical insights into the viable mechanism for the NC operation, and points to a simple yet effective material scheme for achieving hysteresis-free steep slope transistors with reduced fabrication complexity.

## Results

### Characterizations of polycrystalline PZT thin films

We work with 300 nm thick polycrystalline $PbZr_{0.35}Ti_{0.65}O_3$ films deposited on $Pt/Ti/SiO_2/Si$ substrates (see Supplementary Note 1 for deposition details). Figure 1a shows the x-ray diffraction spectrum of a PZT film, which reveals predominant (001) and (111) growth with a small fraction of (110) grains. We estimate the average crystallite size from the full-width-half-maximum ($L$) of the Bragg peaks using the Scherrer Equation, $B(2\theta) = \frac{K\lambda}{L\cos\theta}$, where $K = 1$ is the Scherrer constant, $\theta$ is the Bragg angle, and $\lambda = 1.5406$ Å. The average grain sizes are 27.8 ± 0.8 nm and 27.7 ± 0.5 nm for the (001) and (111) oriented grains, respectively. Atomic force microscopy (AFM) measurements show that these films possess smooth surface morphology (Fig. 1b), with a typical root mean squared roughness of 1-2 nm.

We have characterized the distribution and orientation of the PZT polarization using piezoresponse force microscopy (PFM). Figures 1c-f show the PFM measurements conducted in both vertical (V-PFM) and lateral (L-PFM) modes on the same region of a PZT film. We observed domains with up to 180° phase contrast (Fig. 1c, e) and large amplitude variations (Fig. 1d, f) in both V-PFM and L-PFM, indicating a wide distribution of polarization orientation in the as-grown state of the film. The domains range in size from 20 nm to more than 100 nm, pointing to the presence of a high density of DWs. There is no clear correlation between the domain distribution and the surface morphology, suggesting that the domain formation is not confined by the grain boundaries, which is consistent with previous phase-field simulation[31] and transmission electron microscopy (TEM) studies[32].

Figure 1g shows the polarization $P$ vs. bias voltage ($V_{bias}$) measured in a capacitance structure, which exhibits robust switching hysteresis with remanent polarization of about 0.3 C $m^{-2}$ and coercive voltages of +1.3 V and -1.1 V. The hysteresis becomes negligibly small at the small bias voltage range of ±0.5 V (Fig. 1h insert). Within the hysteresis-free regime, we extracted a dielectric constant of 630−650, which is one to two orders of magnitude higher than those of conventional dielectrics such as $SiO_2$ and $HfO_2$. The dielectric constant shows little variation in this $V_{bias}$ range, and can yield highly efficient doping in the 2D channel[30].

**Steep slope switching in PZT-gated $MoS_2$ transistors**

We mechanically exfoliate few-layer and bilayer $MoS_2$ flakes on PZT and fabricate them into FET devices back-gated by PZT (Fig. 2a, Methods). Figure 2b shows the AFM topography image of a five-layer $MoS_2$ device (Device FL D1, Methods), which conforms well with the PZT surface morphology. We first investigate the transfer characteristic of the device ($I_d$ vs. $V_g$) within the hysteresis-free regime at 300 K. Figure 2c shows $I_d$ vs. $V_g$ taken at source-drain voltage $V_d$ = 0.1 V. For systematic comparison, $I_d$ is scaled by the channel width $W$. Within an ultra-low voltage window $\Delta V_g$ of 0.76 V (-0.26 V to +0.5 V), the device exhibits a high current switching ratio ($I_{on}/I_{off}$) of about $8\times10^6$ in the forward $V_g$-sweep, which clearly reflects the high doping efficiency of the PZT gate.

Figure 2d shows the transfer curves of the device from 290 K to 320 K after pyroelectric correction (Supplementary Note 6). From the quasi-linear regime of the $I_d - V_g$ curves, we extract the field effect mobility $\mu_{\mathrm{FE}} = \frac{1}{C_{\mathrm{PZT}}}\frac{dG}{dV_{\mathrm{g}}}$, where $C_{\mathrm{PZT}}$ is the areal capacitance for 300 nm PZT, $G = LI_{\mathrm{d}}/V_{\mathrm{d}}$ is the 2D conductivity of the channel, and $L$ is the channel length. At 300 K, $\mu_{\mathrm{FE}} = 7.8\ \mathrm{cm^2V^{-1}s^{-1}}$, comparable with previously reported values for $MoS_2$ FETs interfaced with ferroelectrics[25, 27, 28]. In this temperature range, $\mu_{\mathrm{FE}}$ decreases with increasing temperature, following a power law $T$-dependence of $\sim T^{-1.9}$ (Fig. 2e), which can be attributed to phonon scattering[22]. The exponent $\alpha = 1.9$ is between the theoretically predicted values of for single-layer $MoS_2$ ($\alpha = 1.52$)[33] and bulk $MoS_2$ crystals ($\alpha = 2.6$)[34].

Figures 2f-g show the point-by-point SS of the $MoS_2$ FET calculated from the inverse slope of the transfer curves ($\partial V_{\mathrm{g}}/\partial \log(I_{\mathrm{d}})$) in Fig. 2c (Method and Supplementary Note 7). For the forward $V_{\mathrm{g}}$-sweep, we have achieved a minimum subthreshold swing of $SS_{min} \approx 37$ mV dec$^{-1}$ as the device starts to turn on at $I_{\mathrm{d}} \approx 2.5 \times 10^{-14}$ A μm$^{-1}$. The point-by-point SS remains below 60 mV dec$^{-1}$ for over three decades of channel current, getting close to the thermal limit as $I_d$ exceeds $10^{-11}$ A μm$^{-1}$, while a substantial increase in SS occurs for $I_d > 10^{-10}$ A μm$^{-1}$. The average subthreshold swing in this current range ($10^{-13} - 10^{-10}$ A μm$^{-1}$) is $\mathrm{SS_{avg}} \approx 57 \pm 1$ mV dec$^{-1}$. In the reverse scan (Fig. 2g), SS is close to and fluctuating around 60 mV dec$^{-1}$ in the channel current range of $10^{-11} - 10^{-10}$ A μm$^{-1}$, with $\mathrm{SS_{avg}} \approx 60 \pm 1$ mV dec$^{-1}$. Once $I_{\mathrm{d}}$ exceeds $10^{-12}$ A μm$^{-1}$, the transfer curve of the device is essentially free of hysteresis between the forward and reverse $V_{\mathrm{g}}$-sweeps, agreeing well with the dielectric measurement of the PZT gate (Fig. 1h inset).

Similar switching characteristics have been observed in a bilayer $MoS_2$ device (Device 2L D2). As shown in Fig. 3a, a current switching ratio of $5 \times 10^6$ is achieved in the device within a small voltage window $\Delta V_g$ of 0.76 V (-0.26 V to 0.5 V) in the forward $V_{\mathrm{g}}$-sweep. Compared with the few-layer device, the bilayer channel exhibits a much steeper slope in the initial turn-on characteristic at low channel current, with an $SS_{min}$ of 9.7 mV dec$^{-1}$ at $I_d \approx 8 \times 10^{-14}$ A μm$^{-1}$ (Fig. 3a lower insert). The SS then increases quickly with $I_d$, reaching about 60 mV dec$^{-1}$ at $I_d \approx 10^{-12}$ A μm$^{-1}$. This is in sharp contrast to the moderate $V_g$-dependence of SS observed in the few-layer device at this current range (Fig. 2f and Supplementary Figure 9d). The tradeoff between the steepness of the initial turn-on behavior and the current range of low SS value can be attributed to the competing effects of the channel capacitance $C_{MoS_2}$ and 2D doping efficiency. As shown in Equation (1), $C_{ch}$ not only plays a critical role in stabilizing the NC mode, but also tailors the fractional weight of the second term. As $C_{MoS_2}$ scales inversely with the dielectric layer thickness, the bilayer $MoS_2$ possesses a larger capacitance in the depletion state compared with the few-layer device. For a given $C_g$, it yields a larger fractional weight for the NC term in Equation (1), thus significantly reducing the initial SS value. On the other hand, a thinner channel also corresponds to a lower 2D density of states, requiring a lower $V_g$ to shift the Fermi energy close to the conduction band edge. Once the device reaches the on state, or $C_{MoS_2}$ exceeds $C_{PZT}$, the quasi-static NC mode is no longer energetically favorable, and SS of the device returns to the classical operation regime. Due to the smaller current range where the device exhibits sub-60 mV dec$^{-1}$ SS, the reverse scan exhibits much higher SS that well exceeds 60 mV dec$^{-1}$ (Fig. 3a lower insert), even though the current floor is similar to the few-layer device.

The steep slope switching in the forward $V_g$-sweep is a robust effect observed in multiple PZT-gated $MoS_2$ FETs (Fig. 3b, c and Supplementary Note 8). Figure 3d summarizes the results obtained in devices with the turn-on voltage within the scan range of $\Delta V_g = \pm 0.5$ V. For consistency, we plot the $SS_{avg}$ values averaged over the lowest two decades of $I_d$ (about $10^{-14} - 10^{-12}$ A μm$^{-1}$). For both few-layer and bilayer devices, $SS_{avg}$ is consistently below the classical thermal limit of $k_B T \ln 10/q$ (Equation (1)) over the entire temperature range investigated. Figure 3e compares the $I_{on}/I_{off}$ vs. $\Delta V_g$ result obtained on the few-layer device in Fig. 2 (FL D1) with previous reports for $MoS_2$-based NC-FETs[8-14, 18] and classical FeFETs[24, 25, 27], which highlights the superb performance combination of ultra-low supply voltage and high current on/off ratio in our devices. Despite the large current on-off ratio obtained within a small $\Delta V_g$, these devices exhibit negligible hysteresis ($\delta V_g$ < 10 mV) at $I_d > 10^{-12}$ A μm$^{-1}$. This is in sharp contrast to the widely observed NC-FETs operating upon polarization switching[6, 7, 9, 14, 16, 18], where the hysteresis window between the positive and negative coercive voltages host both on and off states, leading to switching history-dependence that effectively increases the turn-on voltage span and complicates the operation. Between 220 K and 300 K, $\delta V_g$ does not show appreciable variation (Supplementary Figure 11c), showing that the hysteresis-free behavior is not due to the cancellation of polarization switching hysteresis and interfacial charge induced anti-hysteresis, as these two mechanisms have different temperature dependences[35]. Despite the simple material scheme, the minimum SS of 9.7 mV dec$^{-1}$ observed in our bilayer $MoS_2$ device is comparable with the best result reported in hysteresis-free NC-FETs (5.6 mV dec$^{-1}$ in Ref. [13]) using ferroelectric/dielectric stack gates.

We have also characterized the transfer characteristics at different $V_d$ values (Fig. 3b) and scan rates (Fig. 3c). While the device exhibits consistent steep slope switching at different $V_d$ values (Fig. 3b), the transfer curve shifts slightly with increasing $V_d$, with the direction depending on the $V_g$-range (Supplementary Figure 10a, b), which may result from the (negative-)drain-induced-barrier-lowering effect, or (N-)DIBL. The N-DIBL and the associated negative differential resistance is due to the fact that the $MoS_2$ FET is more accurately described as a junctionless transistor[36]. The change of the $V_g$-shift direction can be attributed to the Schottky junction forming between the $MoS_2$ and Ti/Au contact[21, 29], which dominates the channel conduction when the Fermi energy is deep in the gap. For the forward scan, the transfer curve does not show apparent dependence on the scan rate below 20 mV $s^{-1}$ (Supplementary Figure 11a) even in the steep slope switching regime, confirming that the NC effect is intrinsic to the change of the polar state in the polycrystalline PZT gate. At 20 mV $s^{-1}$, the data acquisition speed cannot keep up with the fast scan rate, which leads to a shift in the SS vs. $I_d$ curve and few data points in the sub-60 mV $dec^{-1}$ regime (Fig. 3c). For the reverse scan, the saturation current level increases quasi-linearly with the scan rate (Supplementary Figure 11a), suggesting that the higher current floor in the reverse scan is induced by extrinsic mechanisms. Similar charge hysteresis has been widely observed in 2D FETs with oxide back-gates[13, 16, 30, 35]. The likely culprits include the dynamic response of interfacial adsorbates (*e.g.*, dissociation/recombination of water molecules) and charge trapping/de-trapping. These processes become activated above a threshold bias and respond actively to the varying displacement field, thus screening the 2D channel from the field effect modulation and resulting in a saturated off-current level in the reverse scan[30, 35].

Another possible mechanism for the higher current floor in the reverse scan is the leakage current $I_{leak}$ associated with the high densities of DWs and grain boundaries in the polycrystalline PZT gate. Previous conductive probe AFM (c-AFM) studies have revealed thermally activated, diode-like conduction at DWs in PZT[37]. Unlike the sparse distribution of DWs in single crystalline epitaxial thin films[37, 38], in our studies, the DWs exist within the nanoscale crystallites in the polycrystalline PZT film[32]. As a result, the DWs and grain boundaries form a complex three-dimensional (3D) network of conduction paths rather than a well-defined direct conducting channel between the top and bottom electrodes. This is consistent with our c-AFM studies, which reveal a uniform distribution of leakage current with no direct correlation to the surface grain/domain distributions (Supplementary Figure 7). High precision measurements of the leakage current further confirm that $I_{leak}$ remains below $I_d$ over the entire $V_g$-sweep range (Supplementary Figure 9c). We also note that for the transfer characteristics measured at a larger $V_g$-sweep range (±1 V), which approaches the coercive voltage, current spikes due to partial polarization switching is clearly visible in the leakage current (Supplementary Figure 10d). The transfer characteristics and corresponding SS, on the other hand, remain qualitatively similar (Supplementary Figure 10b, c), further ruling out the impact of $I_{leak}$ on the measured transfer characteristics.

**Theoretical modeling of the DW enabled NC effect**

As shown in Equation (1), the key to accessing the NC mode is to have the gate-induced surface potential change in the semiconductor channel exceed the applied voltage, or $\frac{\partial V_g}{\partial \psi_s} < 1$. This requires the second-order derivative of the Gibbs free energy to be negative, which can be realized in ferroelectrics below $T_C$ near the centrosymmetric transition state during polarization reversal, as shown in Fig. 4a. The initial proposal of the NC-FET device concept thus builds on this polarization switching regime[2]. Close to $E_c$, a relatively small change in $V_g$ can cause polarization reversal by going through the NC state, resulting in a sudden boost in polarization, surface potential $\psi_s$, and surface charge density $Q$ in the semiconducting channel, and hence $I_d$. Figure 4b illustrates how a polarization reversal in PZT ($P = 0.5$ C m$^{-2}$) can cause a jump in the surface potential by $\Delta\psi_s$ = 1.54 V, assuming $Q \approx P$ (Supplementary Note 10). As polarization reversal is a first-order process, it inevitably leads to switching hysteresis. The presence of a dielectric layer, however, can suppress the double well energy. When the potential barrier becomes comparable to the thermal energy, the hysteresis can be quenched.

Unlike previous experimental studies of NC-FETs based on ferroelectric/dielectric stack gates[6-14], the sub-60 mV dec$^{-1}$ SS acquired in our devices in the hysteresis-free regime of PZT suggests the existence of a quasi-static NC mode in absence of an additional dielectric layer and hence the associated capacitance matching. The SS falls below the Boltzmann limit at an applied field well below the coercive field ($E_c$) of the ferroelectric gate, further suggesting that it is not driven by polarization switching. Besides polarization reversal, it has been theoretically predicted that a sudden change of $\psi_s$ in the semiconductor channel can also be achieved through ferroelectric polarization rotation from the in-plane to out-of-plane orientation, which can lead to hysteresis-free operation with higher speed and lower energy consumption[17]. To identify the origin of the observed steep slope switching, we have performed a series of PFM imaging on PZT with different DC bias voltages applied to the sample (Supplementary Notes 2-3). The V-PFM measurements taken at $V_{bias}$ of -0.5 to 0.5 V show that the domain distribution for the out-of-plane polarization remains qualitatively intact (Supplementary Figure 2), consistent with the hysteresis-free dielectric response within this $V_g$-range (Fig. 1h inset). We thus rule out partial ferroelectric switching as a dominant mechanism for the steep slope switching behavior. Similar results have been observed in the L-PFM imaging (Supplementary Figure 3), suggesting that polarization rotation is also not occurring on the large scale in this bias range.

Given the multi-domain nature of the polycrystalline PZT film, we next consider the possible contribution of the ferroelectric DW, a region where the polarization is frustrated[39]. To understand the role of domain formation, we carried out 3D force field simulations based on the Landau-Ginzburg-Devonshire (LGD) theory[40], in which the thermodynamic potential (Gibbs free energy) $F$ can be expressed as:

$$F = \int_V \left(f_{\text{bulk}} + f_{\text{elas}} + f_{\text{grad}} + f_{\text{elec}}\right) dV \, . \tag{2}$$

Here $f_{\text{bulk}}$, $f_{\text{elas}}$, $f_{\text{grad}}$, and $f_{\text{elec}}$ are the energy densities associated with the thermodynamic potential of a PZT single crystal, elastic energy, dipole gradient, and electrostatic energy, respectively (see Supplementary Note 10 for modeling details). Figure 4c shows the simulation result for an equilibrated multiple-domain structure in PZT. It includes equal volumes of up and down polarization domains, which minimizes the electrostatic energy ($f_{\mathrm{elec}}$) cost induced by depolarization field[41]. Near the surfaces, the dipoles mostly lie in the plane to satisfy the continuity of electric displacement and minimize $f_{\mathrm{grad}}$, leading to flux-closure type chiral dipole structures at the DWs. The existence of this type of chiral dipole structure has been demonstrated experimentally in ferroelectric systems with enhanced depolarization field, including the surface polar rotation at ferroelectric DWs[42, 43], the structural phase boundaries,[44] and the polar vortices in ferroelectric/dielectric superlattices[15]. Monte Carlo simulations have shown that these polar structures can account for the NC-type dielectric response observed in the latter system[45]. In polycrystalline PZT films, as the crystallites are small in size and have high surface to bulk ratios, it is natural to expect high depolarization field and abundant DWs[32]. Theoretical modeling of polycrystalline ferroelectrics has also revealed the appearing of chiral polar rotation at the 90° DWs as well as at the 180° DWs close to the grain boundaries[31].

We then impose a gate voltage $V_{\mathrm{g}}$ on this equilibrated model in a sweeping sequence of 0 → 0.5 → 0 → -0.5 → 0 V and calculate the evolution of local dipoles (Fig. 4c) and associated channel current with respect to $V_{\mathrm{g}}$. Figure 4d shows the simulated profile for the polarization change ($\Delta P$) upon sweeping $V_{\mathrm{g}}$ across -0.25 V, where an abrupt increase in the polarization occurs only at the DWs. This local boost of $\Delta P$ can be well correlated with a sudden jump in $I_{\mathrm{d}}$, as shown in the simulated transfer curves (Fig. 4e). The corresponding SS reaches the minimum value of about 13.6 mV dec$^{-1}$ at $I_{\mathrm{d}} \approx 2.6 \times 10^{-13}$ A μm$^{-1}$ and remains below the 60 mV dec$^{-1}$ limit till $I_{\mathrm{d}}$ reaches $\sim 10^{-10}$ A μm$^{-1}$. Without considering the current floor imposed by the extrinsic charge contribution, the forward and backward gate sweeps overlap with each other. We next introduce an extrinsic current contribution $I_0$ into the model (Supplementary Note 10), which can be due to either the interfacial charge dynamics or measurement noise. It successfully reproduces the softening of the turn-on behavior of the transfer curve (blue curve in Fig. 4e and insert). The simulated transfer characteristics thus well capture the main features of the experimental observation.

**Correlating steep slope switching with DW density**

To establish the relationship between the DW and the steep slope switching, we have controlled the DW density in PZT by applying a negative DC bias voltage that is sufficiently high to trigger polarization rotation to be correlated with the transfer characteristics of $MoS_2$ FETs at the corresponding $V_g$-sweep range. Figure 5a shows a series of L-PFM images taken on the same area of a PZT film with progressively higher DC $|V_{bias}|$ applied during scan (Supplementary Note 3). While the domain distribution at $V_{bias}$ = -0.5 V is essentially the same as that of zero bias, the sample gradually approaches a more uniform polar state as $V_{bias}$ exceeds the coercive voltage. We then identify the DW positions in the L-PFM phase image (Fig. 5a), and extract the DW length per unit area $l_{DW}/A$ (Supplementary Note 4). At $V_{bias}$ = 0 V, $l_{DW}/A$ = 0.039±0.002 $nm^{-1}$, comparable with that obtained from Fig. 1e (0.036±0.002 $nm^{-1}$). This value is also in good agreement with previous TEM studies of domain formation in polycrystalline PZT[32], suggesting a similar level of DW density at the sample surface and within the grain inside the samples. Based on TEM imaging of the PZT DW close to surface, we estimate that the width of the chiral polar state within DW is about 3 nm [42]. This yields a DW areal density of 11.8% for the surface layer of grains in PZT. Note this is an underestimate, as L-PFM is not sensitive to DWs along the scan directions. As shown in Fig. 5b, $l_{DW}/A$ decreases monotonically with increasing $|V_{bias}|$.

We then investigate the transfer curves of multiple 2L $MoS_2$ FETs fabricated on a PZT film, having them characterized at different $V_g$-sweep ranges. At $|V_g| \geq 1.5$ V, the $I_d$ vs. $V_g$ curves exhibit an anti-hysteresis, similar to that observed on the device back-gated by $SiO_2$ (Supplementary Figure 12), which can be attributed to the charge dynamics of interfacial adsorbates and/or defect states[30, 35]. Regardless the high-$V_g$ hysteresis, steep slope switching has been observed in all forward scans. The $SS_{min}$ value increases monotonically with increasing $V_g$-sweep range, which correlates well with the reduced DW areal density (Fig. 5b and Supplementary Note 3). For $V_g$-sweep range of ±2 V, which exceeds the coercive voltage, the $SS_{min}$ of the device is approaching the 60 mV $dec^{-1}$ thermal limit, further ruling out polarization reversal as the origin of the NC effect. This result thus yields strong support to the scenario of DW enabled NC modes in the $MoS_2$ FETs.

## Discussion

With the simulation results and PFM studies, we attribute the experimentally observed steep slope switching to the NC states of the DWs, which are abundant in polycrystalline PZT (Fig. 1c-f)[31, 32]. In this scenario, the DW region possesses significantly suppressed local polarization due to their neighbors with antiparallel dipole orientations. These metastable polar states are delicate and have the tendency to collapse into a uniform polarization upon external perturbation, as evidenced by the enhanced dielectric susceptibility observed in the DWs[15, 32, 45]. An external electric field can thus induce a much larger increase in the dipoles at the DW compared with the dipoles inside the uniformly polarized domains. The simulated jump in the polarization is about 0.004 C $m^{-2}$ (Fig. 4d), which is on the same order of magnitude as that measured in our PZT films at $V_{bias}$ below the coercive voltage (Fig. 1h inset). Even though this polarization value is much smaller than the remanent polarization of bulk PZT, it is comparable with that of polymorphous $(Hf,Zr)O_2$[13] and large enough to induce a significant boost in $\psi_s$ (0.76 V), as indicated in Fig. 4b.

Our proposal of DW induced NC effect does not require capacitance matching from a dielectric layer, which is distinct from the extensively studied scenario operating in the polarization reversal regime. The latter effect capitalizes on the strong depolarization field provided by a dielectric layer to suppress the energy barrier in the ferroelectric between two polarization states, which can stabilize the steady-state NC effect in the hysteresis-free mode[46]. For the DW region, in sharp contrast, the local dipoles are naturally suppressed in a metastable state and thus highly susceptible to external perturbation. Force field simulation shows that the DW region hosts a series of local minima with low energy barrier (Supplementary Figure 13), which strongly resembles that of the ferroelectric/dielectric stack capacitors close to the capacitance matching condition[46]. The energy gap for DW motion in a one-dimensional (1D) dipole chain is orders of magnitude lower than the electrostatic energy for polarization reversal of the uniformly polarized 1D domain (Supplementary Note 11). At low bias field, the DW motion around equilibrium state is still within the thermal activation but well-below the depinning regime[47], which can result in a quasi-linear dielectric response[48], leading to an approximately zero hysteresis window in the $I_d$ vs. $V_g$ curve. This scenario is consistent with the essentially hysteresis-free $P$-$V$ loop within this bias range (Fig. 1h inset) and previous dielectric studies of polycrystalline PZT films with similar thickness[32].

The DW-enabled NC mode is generally applicable to DW-rich ferroelectric systems, such as polycrystalline thin films and films deposited along a crystalline orientation off the major polar axis. For example, steep slope switching has been observed in $MoS_2$ FETs gated by a single layer of polycrystalline P(VDF-TrFE)[14, 18], which may share the same origin. It is worth noting that utilizing the DW enabled NC state to construct the steep slope FETs has distinct advantages in terms of device performance compared with the extensively studied mechanisms based on polarization switching. In conventional NC-FET based on ferroelectric/dielectric stack gate, the NC mode occurs close to or exceeds the coercive field. Domain formation is not desired as it can change the overall energy profile of the composite system, *i.e.*, perturbing the negative curvature and therefore destabilizing the NC state[39, 49]. The DW enabled NC effect, in contrast, capitalizes on the continuity of polarization through different domains[31] and is thus intrinsically hysteresis-free and low energy. As the process does not involve dipole reorientation, it also promises high-speed (GHz) operation[50]. The fact that it does not require an additional dielectric layer further reduces the fabrication complexity.

Similar to the NC-FETs exploiting the ferroelectric/dielectric stack gates, whether the polycrystalline gate can render hysteresis-free steep slope switching depends on the specific material parameters, such as the crystallite orientation and grain size/DW density. As we are working with a 3D network of DWs, the size scaling limit on the lateral device dimension depends on the intricate relation between the polycrystalline grain size and the film thickness of PZT. Due to the high dielectric constant of the polycrystalline PZT films ($\kappa$~650), a 300 nm PZT film only corresponds to an equivalent oxide thickness of 1.8 nm. It can host a considerable number of grains/DWs along the vertical direction, which collectively contribute to the NC effect. From the PFM image, we estimate that the areal density of the DW in a single layer of grains can exceed 11% (Supplementary Note 4)[42]. Given the thickness and average grain size of our PZT films, the number of grains along the film normal is on the order of 10. This means that the actual in-plane DW area can potentially cover the majority of the channel area. The minimum film thickness is thus determined by the critical areal DW density for anchoring the NC effect. Increasing the thickness of the high-$\kappa$ layer, on the other hand, can also increase the stray drain field and eventually lead to extreme short channel effect[51]. The optimal thickness of PZT thus depends on the tradeoff between these two competing effects. As the domain dimension typically scales with the grain size[52], the optimal ferroelectric thickness can be further reduced by working with polycrystalline films with intrinsically small grain size. Future experimental and theoretical studies are needed to map out the material parameter space for achieving the DW-enabled NC effect in polycrystalline ferroelectric films approaching the 10 nm scale. Another key attribute that affects the device performance is DW vibration. In ferroelectrics with low DW stiffness, such as hexagonal rare-earth manganites, it can lead to enhanced dielectric loss at frequencies above 100 MHz[47]. Ferroelectrics with stiff DWs, such as PZT[53], are thus more suitable for high-frequency applications. In addition, controlling the extrinsic interfacial charge condition is critical for achieving hysteresis-free switching at the low current ($I_d < 10^{-12}$ A μm$^{-1}$) regime.

In terms of the $MoS_2$ channel, even though we have achieved similar current on/off ratios in the few-layer and bilayer devices, they exhibit distinct turn-on behaviors. The few-layer $MoS_2$ shows steep-slope switching over three decades of channel current, with only moderate $V_g$-dependence of SS. The bilayer device, in sharp contrast, possesses a much steeper initial turn on with an $SS_{min}$ that is only 26% of the value for the few-layer device, while SS increases rapidly with increasing current. The stabilization of the NC mode depends on the relative length scales of channel thickness and screening length of $MoS_2$. The optimal thickness for $MoS_2$ is contingent upon the channel mobility and required operation current level for the specific applications. The modeling of channel current in Fig. 4e is based on experimentally extracted mobility value of our $MoS_2$ samples (~8 cm$^2$ V$^{-1}$ s$^{-1}$), which limits the on-current level to be $10^{-7}$-$10^{-6}$ A μm$^{-1}$, comparable with the state-of-the-art results for back-gated $MoS_2$ devices[13]. The $MoS_2$ channel mobility can be improved by capping the device with a top *h*-BN or high-$\kappa$ dielectric[22], which would also extend the voltage sweep range for the steep slope switching and reduce the initial SS value.

In summary, we report the experimental demonstration of a prototype DW-enabled 2D steep slope transistor utilizing a single-layer polycrystalline PZT gate without additional dielectric matching. Compared to prior works on PZT-based $MoS_2$ FeFETs and NC-FETs, our devices exhibit comparable current switching ratio at significantly lower $V_g$ and are essentially hysteresis-free over a wide current range. Theoretical modeling reveals the critical role of the metastable polar states within the DW in stabilizing the NC mode. With solution-processed, easy-to-fabricate polycrystalline ferroelectric thin films, single-layer gate geometry, sub-60 mV $dec^{-1}$ SS, and ultra-low working voltages, our work points to a cost-effective material strategy for developing high-performance low-power 2D nanoelectronics.

**Methods**

*Characterization of PZT thin films.* The structural properties of the polycrystalline PZT films are characterized using a Rigaku SmartLab Diffractometer with Cu $K_\alpha$ radiation ($\lambda$ = 1.5406 Å). The surface and PFM characterizations of the PZT films are carried out on a Bruker Multimode 8 AFM. The V-PFM and L-PFM measurements are conducted using Bruker SCM-PIT and NanoSensors PPP-EFM probes with the drive frequencies close to one of the resonant frequencies. The c-AFM measurements are performed in contact mode with Bruker SCM-PTSI probes. For the dielectric/ferroelectric characterizations, we deposit Pt or Au top electrodes on PZT. The dielectric constant of PZT is extracted from the *C*-*V* measurements conducted with an HP 4291A RF Impedance Analyzer between ±0.5 V at 1 kHz. The dielectric constant and dielectric loss of the PZT films only show pronounced changes above 100 kHz (Supplementary Figure 1). The low-voltage *P*-*V* loops are measured with triangular waves using an aixACCT TF analyzer 2000 between $V_{bias}$ = ±0.5 V at 1 kHz. The high-voltage $P$-$V$ loops are measured between $V_{bias}$ = ±5 V with Precision Premier II Ferroelectric Tester (Radiant Technologies, USA) at 1 kHz.

*Fabrication and characterization of $MoS_2$ devices.* We mechanically exfoliate $MoS_2$ flakes on elastomeric films (Gel-Film® WF×4 1.5mil from Gel-Pak) from bulk single crystals. Few-layer flakes are identified using optical microscopy and Raman spectroscopy and transferred onto PZT. The dimension of the flakes varies from 2 μm to tens of microns (Supplementary Table 1). For the device shown in Fig. 2b, the frequency difference $\Delta\omega$ between the Raman $E_{2g}^{1}$ and $A_{g}^{1}$ modes is 24.4 $cm^{-1}$, corresponding to about five-layer $MoS_2$. For the bilayer device shown in Fig. 3a, $\Delta\omega$ is 21.9 $cm^{-1}$. We then fabricate the $MoS_2$ samples into two-point devices using e-beam lithography followed by evaporation of 5 nm Ti/50 nm Au electrodes. The results shown in this work are based on 3 PZT-gated bilayer $MoS_2$ FETs (denoted as Device 2L D1-D3), 3 PZT-gated few-layer devices (denoted as Device FL D1-D3), and 1 $SiO_2$-gated bilayer $MoS_2$ FET (denoted as Device 2L D4). The physical dimension of the $MoS_2$ FETs was summarized in Table S1 in the Supplementary Information. The variable temperature electrical characterizations of the $MoS_2$ FETs are performed on either the Quantum Design PPMS or the Lakeshore TTP4 probe station. For measurements taken on the PPMS, $I_d$ is measured between the source and drain contacts using Keithley 6430 Sub-Femtoamp Remote SourceMeter, and $V_g$ ($I_{leak}$) is applied (measured) between the gate and drain contacts via a Keithley 2400 SourceMeter. For measurements carried on the probe station, the transfer curves are taken using Keysight 1500A Semiconductor Analyzer, where $I_d$ and $I_{leak}$ are measured via the high precision ports and $V_g$ is applied via the medium precision port. The transfer curves are taken at $V_d$ = 0.1-0.5 V with $V_g$-sweep at a step size of 10 mV. For all devices characterized, the measured instrument current floor is about 100 fA.

The SS vs. $I_d$ curves in Fig. 2 and Fig. 3 are calculated by differentiating $I_d$ at each $V_g$ point on the $I_d$-$V_g$ curves. This approach is widely used for assessing SS in NC-FETs due to their special operation characteristics[6, 9, 13, 14]. Unlike transistors using conventional dielectric gate, the NC effect in a ferroelectric-gate only occurs within a very narrow $V_g$-window, *e.g.*, in the vicinity of either the coercive voltage for polarization switching or the critical voltage for polarizing the DWs, which can be clearly tracked by the point-by-point method. In Supplementary Figure 9d, we show the averaged subthreshold SS of the few-layer and bilayer devices by taking the slope of any three consecutive points. The results are consistent with those obtained using the point-by-point method within the error bar.

**Data Availability statement**

The data that support the plots within this paper and other findings of this study are available from the corresponding author upon reasonable request.

**Acknowledgements**

We would like to thank Patty Niemoth, Tianlin Li, and Pratyush Buragohain for technical assistance, Yongfeng Lu and Alexei Gruverman for providing equipment access, and Zoran Krivokapic for inspiring discussions. This work was primarily supported by the U.S. Department of Energy (DOE), Office of Science, Basic Energy Sciences (BES), under Award No. DE-SC0016153 (scanning probe studies, device fabrication and characterization). Additional support is provided by the National Science Foundation (NSF) Grant No. OIA-1538893 (PZT preparation), Semiconductor Research Corporation (SRC) under GRC Task Number 2831.001 and the Nebraska Center for Energy Sciences Research (PZT characterization). A.M.R. and Y.Q. acknowledge the support from the U.S. Department of Energy (DOE), Office of Science, Basic Energy Sciences (BES), under Award No. DE-FG02-07ER46431 and computational support from the National Energy Research Scientific Computing Center (NERSC) of the DOE (theoretical modeling). The research was performed in part in the Nebraska Nanoscale Facility: National Nanotechnology Coordinated Infrastructure and the Nebraska Center for Materials and Nanoscience, which are supported by NSF under Award ECCS: 2025298, and the Nebraska Research Initiative.

**Author Contributions**

X.H. conceived and supervised the project. S.-H.K. and A.I.K. prepared the PZT films. J.S. and S.-H.K. performed structural and electrical characterizations of the PZT films. J.S. and K.W. conducted the PFM and c-AFM studies. J.S., Z.X., and D.L. fabricated the $MoS_2$ FETs and carried out the Raman and electrical characterizations. Y.Q. and A.M.R. performed the modeling of the $MoS_2$ FETs and DWs. J.S., Y.Q. and X.H. wrote the manuscript. All authors discussed the results and contributed to the manuscript preparation.

**Competing Interests**

The authors declare no conflict of interest.

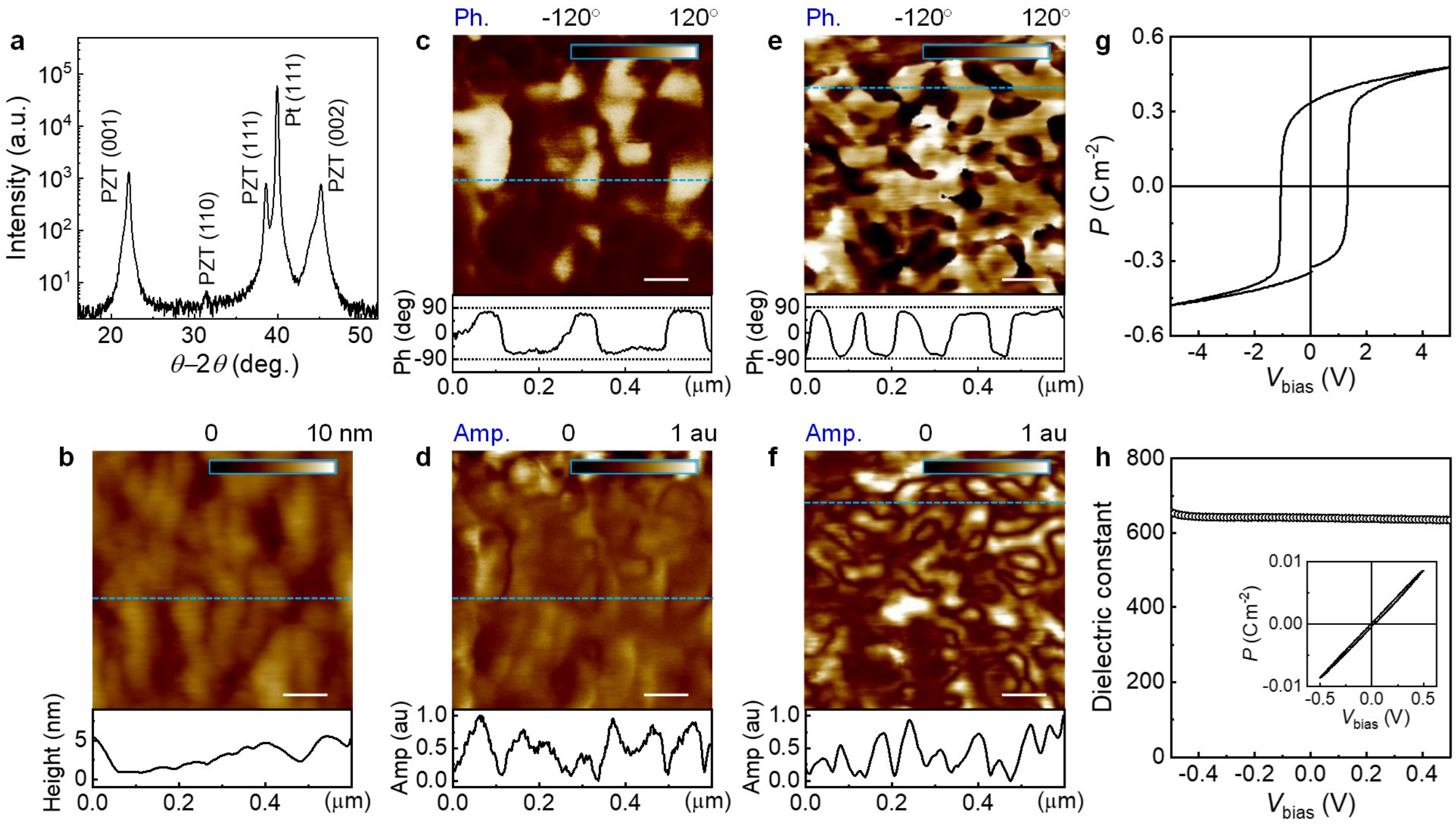

**Fig. 1 Characterization of polycrystalline PZT films. a** X-ray $\theta$-$2\theta$ scan taken on a 300 nm PZT film. **b** AFM topography, and **c-f** PFM images of the same area on a PZT film. **c** V-PFM phase and **d** amplitude images. **e** L-PFM phase and **f** amplitude images. The lower panels show the signal profiles along the dashed lines. Scale bars in **c-f** represent 100 nm. **g** $P$ vs. $V_{\mathrm{bias}}$ hysteresis taken on a PZT film. **h** Dielectric constant of the film vs. $V_{\mathrm{bias}}$, with $V_{\mathrm{bias}}$ well below the coercive voltage. **Inset:** $P$ vs. $V_{\mathrm{bias}}$ taken at this voltage range.

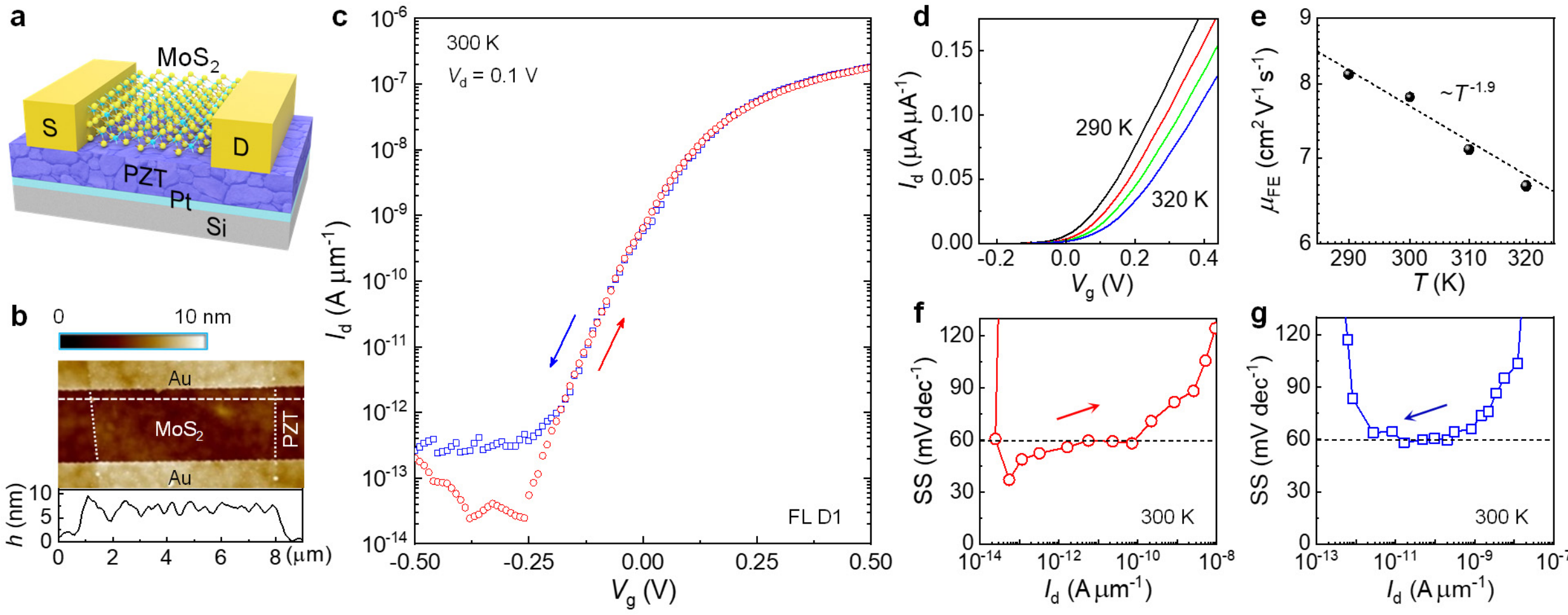

**Fig. 2 Characterization of a few-layer $MoS_2$ FET.** **a** Device schematic. **b** AFM topography image of a few-layer $MoS_2$ device (FL D1) with the height profile along the dashed line (lower panel). The dotted lines outline the $MoS_2$ flake. **c** Transfer characteristics of the $MoS_2$ FET at 300 K in both forward and reverse $V_g$-sweeps at scan rate of 10 mV $s^{-1}$. **d** Transfer characteristics of the device at various temperatures (top to bottom: 290, 300, 310, and 320 K), and **e** the corresponding $\mu_{FE}$ vs. $T$ with a fit to $T^{-1.9}$. **f, g** Point-to-point SS vs. $I_d$ extracted from panel **c** in **f** forward $V_g$-sweep and **g** reverse $V_g$-sweep. The dashed lines depict the thermal limit for SS at 300 K.

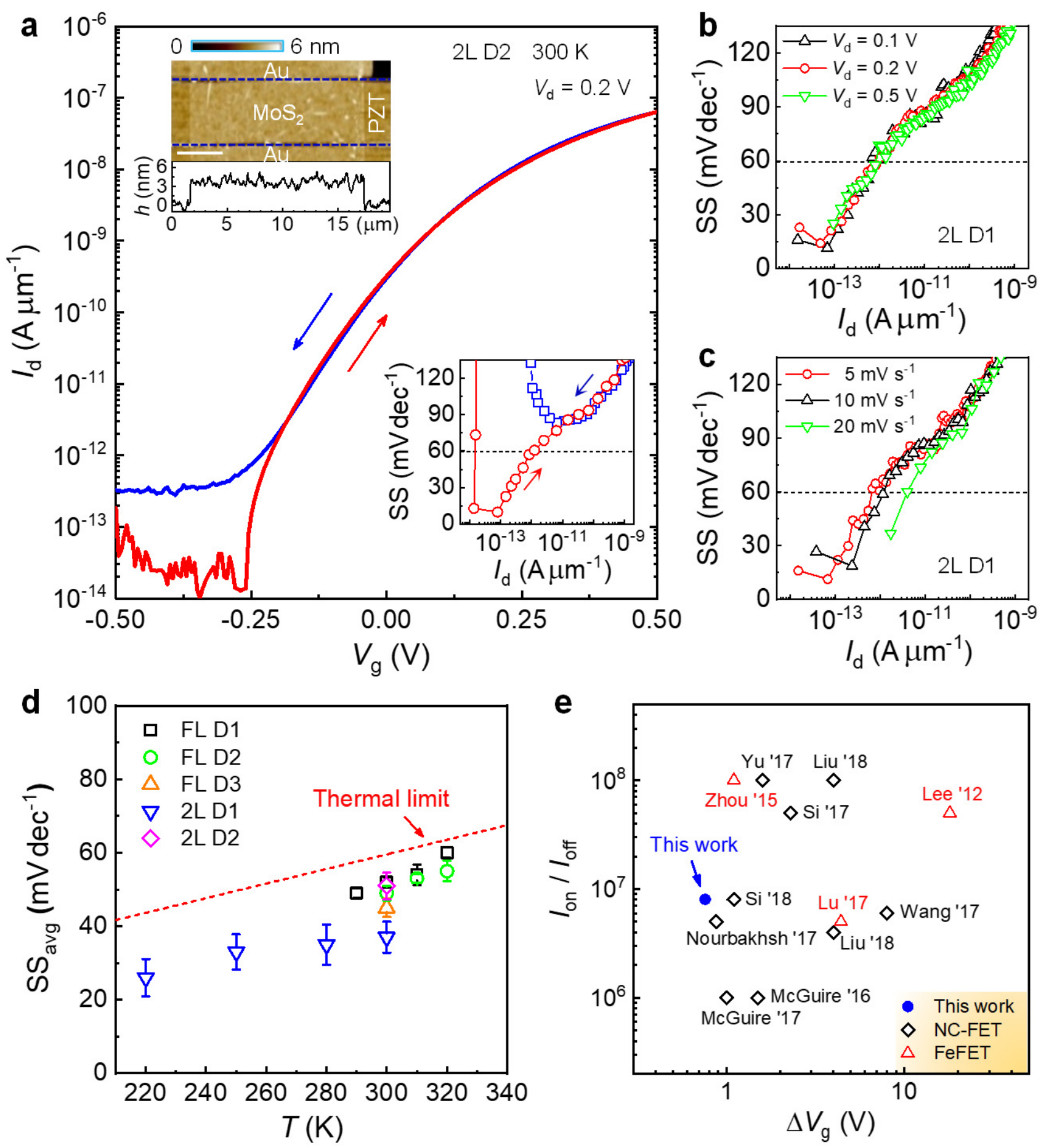

**Fig. 3 Performance of the $MoS_2$ NC-FETs. a** Transfer characteristics of a bilayer $MoS_2$ FET (2L D2) at 300 K in both forward and reverse $V_g$-sweeps at scan rate of 5 mV s$^{-1}$. **Top inset:** AFM topography with the height profile averaged over the entire channel (lower panel). Scale bar represents 4 μm. **Bottom inset:** The corresponding point-to-point SS vs. $I_d$. The dashed line depicts the thermal limit for SS at 300 K. **b, c** Point-to-point SS vs. $I_d$ taken on Device 2L D1 extracted at **b** different $V_d$ and **c** different $V_g$-scan rates. **d** $SS_{avg}$ vs. $T$ taken on three few-layer and two bilayer $MoS_2$ FETs, including the data shown in **Fig. 2** and **Fig. 3a-c**. The red dashed line depicts the theoretical Boltzmann limit of SS. **e** Current on/off ratio vs. required $\Delta V_g$ taken from the current work (solid symbol) and those from literature (open symbols)[8-14, 18, 24, 25, 27].

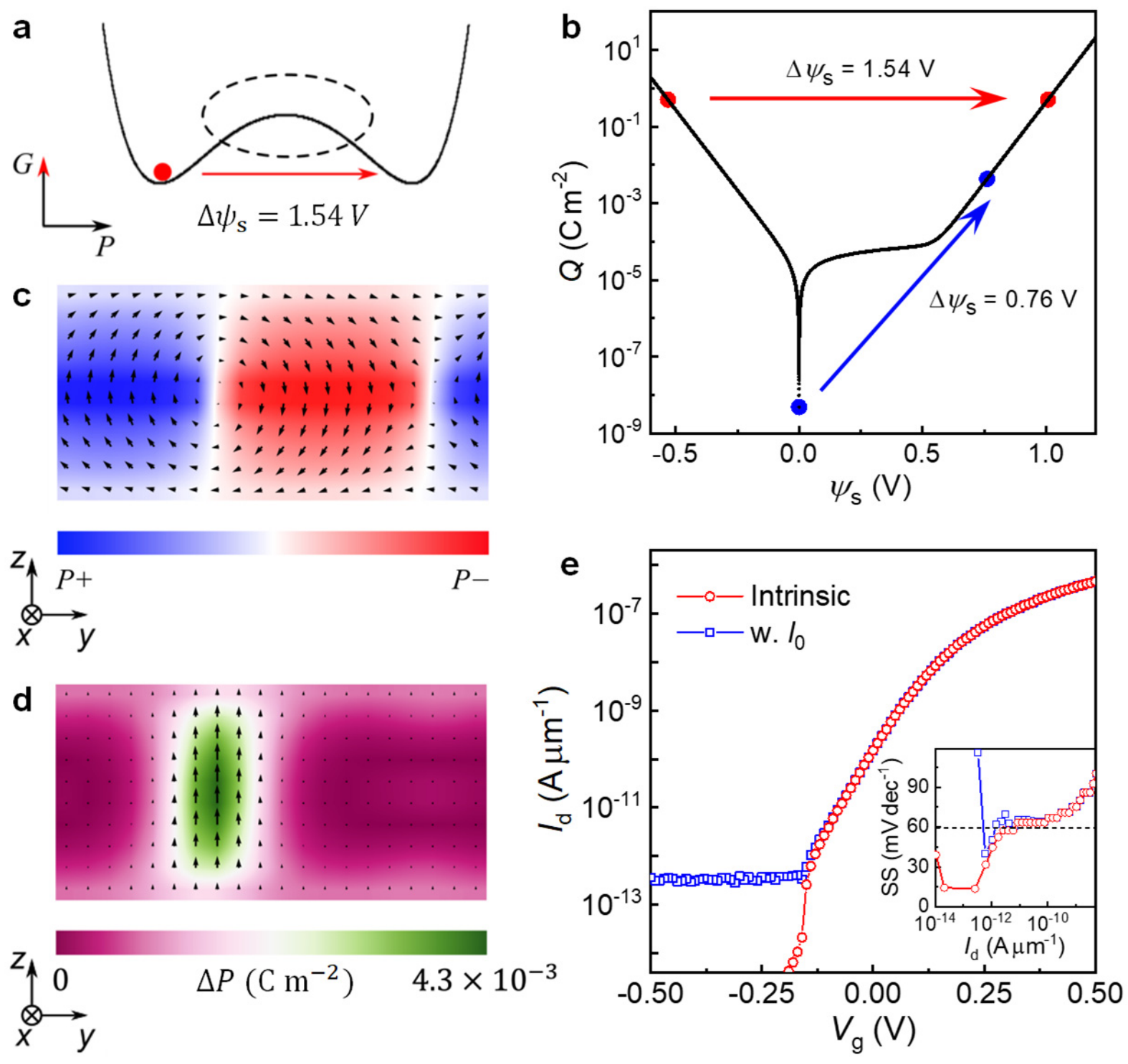

**Fig. 4 3D force field simulation results. a** Schematic double-well Gibbs free energy profile of a ferroelectric material showing the negative capacitance regime (circled). **b** 2D charge density $Q$ vs. $\psi_s$ curve for $MoS_2$. For a ferroelectric gate, the continuity of electric displacement yields $Q \approx P$. The red arrow indicates the change of $\psi_s$ during a polarization flipping (from $Q = -0.5\ \mathrm{C\ m^{-2}}, \psi_s = -0.53\ \mathrm{V}$ to $Q = +0.5\ \mathrm{C\ m^{-2}}, \psi_s = 1.01\ \mathrm{V}$). The blue arrow indicates the change of $\psi_s$ during a polarization increase (from $Q = 0\ \mathrm{C\ m^{-2}}, \psi_s = 0\ \mathrm{V}$ to $Q = +0.0043\ \mathrm{C\ m^{-2}}, \psi_s = 0.76\ \mathrm{V}$). **c** Simulated multiple-domain structure in PZT with two inequivalent DWs. The left DW hosts a polar vortex, and the right one hosts an anti-vortex. **d** Simulated change of polarization after and before the $I_d$ jump around $V_g \approx$ -0.25 V. **e** Simulated $I_d$ vs. $V_g$ curves without (red) and with (blue) an extrinsic current contribution $I_0$ taken into account, and the corresponding SS vs. $I_d$ (inset). The dashed line depicts the thermal limit for SS at 300 K.

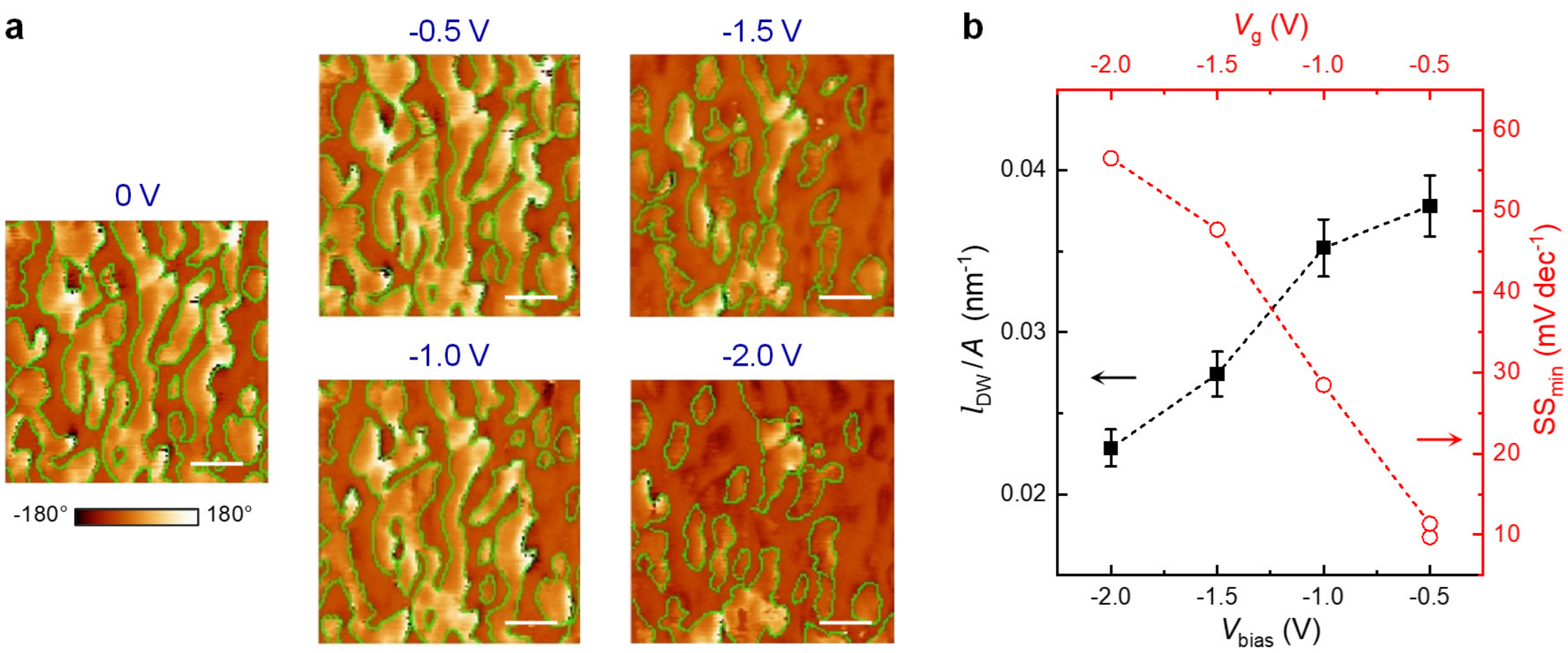

**Fig. 5 Relation between DW density and SS. a** L-PFM phase images of the same area in a 275 nm PZT film taken at progressively higher DC $V_{bias}$. All scale bars are 100 nm. The green lines mark the identified DW positions. Scale bars represent 100 nm. **b** $l_{DW}/A$ vs. $V_{bias}$ extracted from **(a)** (left-bottom axes) and $SS_{min}$ vs. the $V_g$ scan range taken on 2L $MoS_2$ FETs (right-top axes).

# Supplementary Information

## Domain Wall Enabled Steep Slope Switching in $MoS_2$ Transistors Towards Hysteresis-Free Operation

Jingfeng Song[1], Yubo Qi[2], Zhiyong Xiao[1], Kun Wang[1], Dawei Li[1], Seung-Hyun Kim[3], Angus I. Kingon[3], Andrew M. Rappe[2] and Xia Hong[1*]

[1] *Department of Physics and Astronomy & Nebraska Center for Materials and Nanoscience, University of Nebraska-Lincoln, Lincoln, NE 68588, USA*

[2] *Department of Chemistry, University of Pennsylvania, Philadelphia, PA 19104-6323, USA*

[3] *School of Engineering, Brown University, Providence, RI 02912, USA*

[*] email: xia.hong@unl.edu

### Supplementary Note 1: Sample Information

In this study, we work with spin-coated polycrystalline $Pb(Zr_{0.35}Ti_{0.65})O_3$ (PZT) thin films[1]. The PZT solid solution is prepared using Pb-acetate trihydrate [$Pb(CH_3COO)_2•3H_2O$, 99.9% High Purity Chemicals Inc.], Ti-tetra-i-propoxide [$Ti(O\text{-}i\text{-}C_3H_7)_4$, 99.999% High Purity Chemicals Inc.] and Zr-tetra-n-butoxide [$Zr(O\text{-}n\text{-}C_4H_9)_4$, 99.9% High Purity Chemicals Inc.] as main metal precursors. For environment compatibility, we choose non-toxic butanol (>99.4 %, Sigma-Aldrich) as a reactant and solvent instead of the toxic 2-methoxyethanol. For preparation of high-quality solid solutions, we use acetylacetone (>99%, Sigma-Aldrich) as a chelating agent for PZT solution to reduce the reactivity of the hydrolysis and condensation reactions. Besides chelating effects, acetylacetone enhances a dissolving power for the precursors in the solution. We select the Zr/Ti composition to be 35/65. Approximately 15% excess Pb is added to the precursor solution to compensate for the loss of this component during the annealing process. The PZT films are spin coated on Pt(150 nm)/Ti(10nm)/$SiO_2$(300nm)/Si (Quintess Co. Ltd., Korea) at 3000 rpm for 30 seconds, and we subsequently perform a two-step drying process (450 °C for 5 min. and 650 °C for 2 min.) for each layer. After all layers are deposited, the coated films are annealed at 650 °C for 30 minutes in air atmosphere by direct insertion into a heated tube furnace. The thickness of each layer is 100 nm and the final film thickness is 300 nm. For *P-V* measurements, a Pt or Au layer is deposited as top electrode. To reduce the interface damage during metal deposition, the sample is annealed for 10 to 15 min at 600–650 °C for Pt top electrode and 500–550 °C for Au electrode before electrical measurements.

We have conducted *C-V* and *C-F* measurements at 10-100 kHz range and obtained similar results with the 1 kHz measurement. Supplementary Figure 1a shows the bias dependence of the dielectric constant extracted from the *C-V* measurements at 10 kHz, which is consistent with that obtained at 1 kHz shown in Fig. 1h in the main text. Supplementary Figure 1b shows the frequency-dependence of the extracted dielectric constant and dielectric loss, which only exhibits significant variation above 100 kHz. The enhanced dielectric loss above 100 kHz is consistent with the frequency for domain wall (DW) vibration[2].

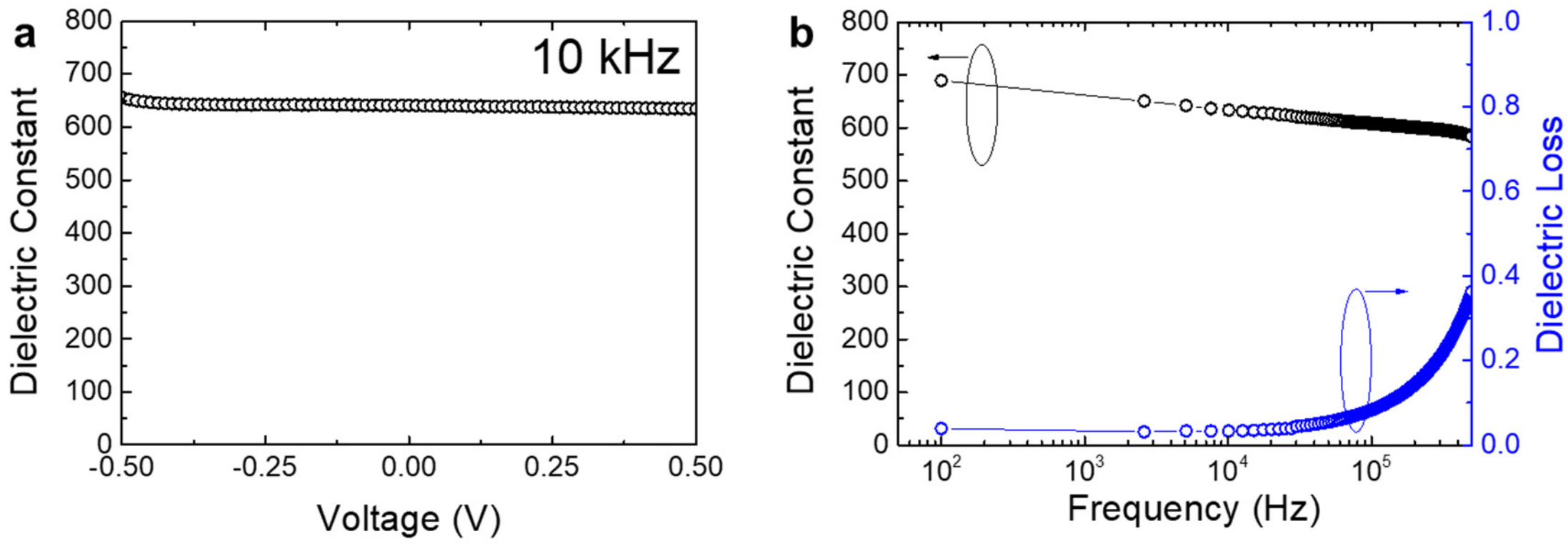

**Supplementary Figure 1 Dielectric measurement of polycrystalline $PbZr_{0.35}Ti_{0.65}O_3$ thin film. a** Dielectric constant of 300 nm polycrystalline $PbZr_{0.35}Ti_{0.65}O_3$ film vs. bias voltage measured at 10 kHz. **b** Dielectric constant (black) and dielectric loss (blue) vs. frequency at zero bias.

Supplementary Table 1 summarizes the physical dimensions of the few-layer (FL) and bilayer (2L) $MoS_2$ FETs studied in this work.

| Sample | Gate | $MoS_2$ Layer # | Channel Width (μm) | Channel Length (μm) |
|---|---|---|---|---|
| FL D1 | PZT | 5±1 | 7.4 | 3.0 |
| FL D2 | PZT | 5±1 | 7.0 | 2.3 |
| FL D3 | PZT | 5±1 | 6.6 | 3.5 |
| 2L D1 | PZT | 2 | 15.7 | 6.0 |
| 2L D2 | PZT | 2 | 15.9 | 5.0 |
| 2L D3 | PZT | 2 | 9.2 | 2.8 |
| 2L D4 | $SiO_2$ | 2 | 8.0 | 3.0 |

**Supplementary Table 1 Physical dimensions of the $MoS_2$ FETs.** The uncertainty in the layer number of few-layer $MoS_2$ originates from the fact that the Raman shift for 4-6 layer $MoS_2$ are very close.

## Supplementary Note 2: PFM Studies of Polycrystalline PZT Films at Low Bias Voltage

To probe the variation of the local domain structure within the applied gate voltage ($V_g$) range for the field effect transistor (FET) devices, we have performed piezoresponse force microscopy (PFM) measurements at different DC bias voltages. Supplementary Figure 2 shows a series of vertical PFM (V-PFM) images taken in the same region on a 275 nm polycrystalline PZT film, with a DC bias ($V_{bias}$) applied across the film during imaging. To be consistent with $V_g$ with respect to the $MoS_2$ channel, $V_{bias}$ varying from -0.5 V to +0.5 V is applied between the bottom electrode and top electrode (AFM tip). The smallest domains resolved are at the 10 nm scale. As this film is slightly thinner than the PZT-gate for the $MoS_2$ FET devices (300 nm), this $V_{bias}$ range can well capture the effect of $V_g$ on the domain evolution in the negative capacitance (NC)-FETs. We find most of the out-of-plane polar domains remain qualitatively similar at negative $V_{bias}$. Consistent results are obtained in multiple samples/locations. We thus rule out partial ferroelectric switching as the mechanism for the steep slope switching behavior. Similar results have also been observed in the lateral PFM (L-PFM) imaging (Supplementary Figure 3), suggesting that polarization rotation is also not occurring on the large scale in this bias range.

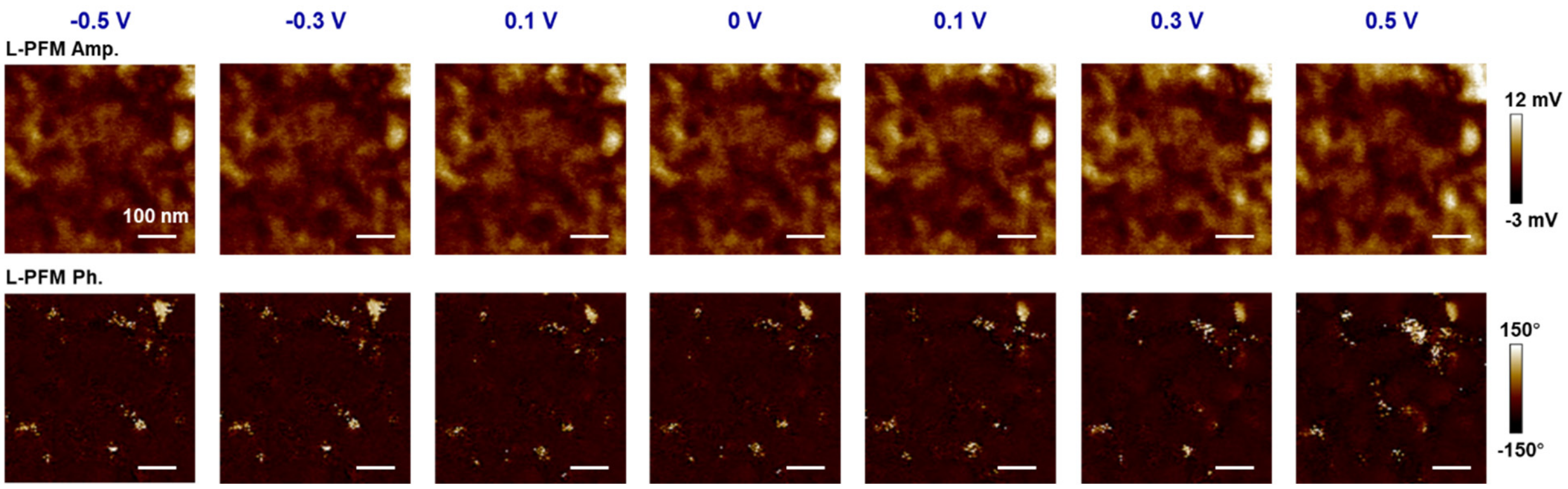

**Supplementary Figure 2 V-PFM imaging of PZT film at low $V_{bias}$.** V-PFM amplitude (top) and phase (bottom) images of the same area in a 275 nm PZT film taken at different DC bias voltages. The AC drive voltage is 100 mV. All scale bars are 100 nm.

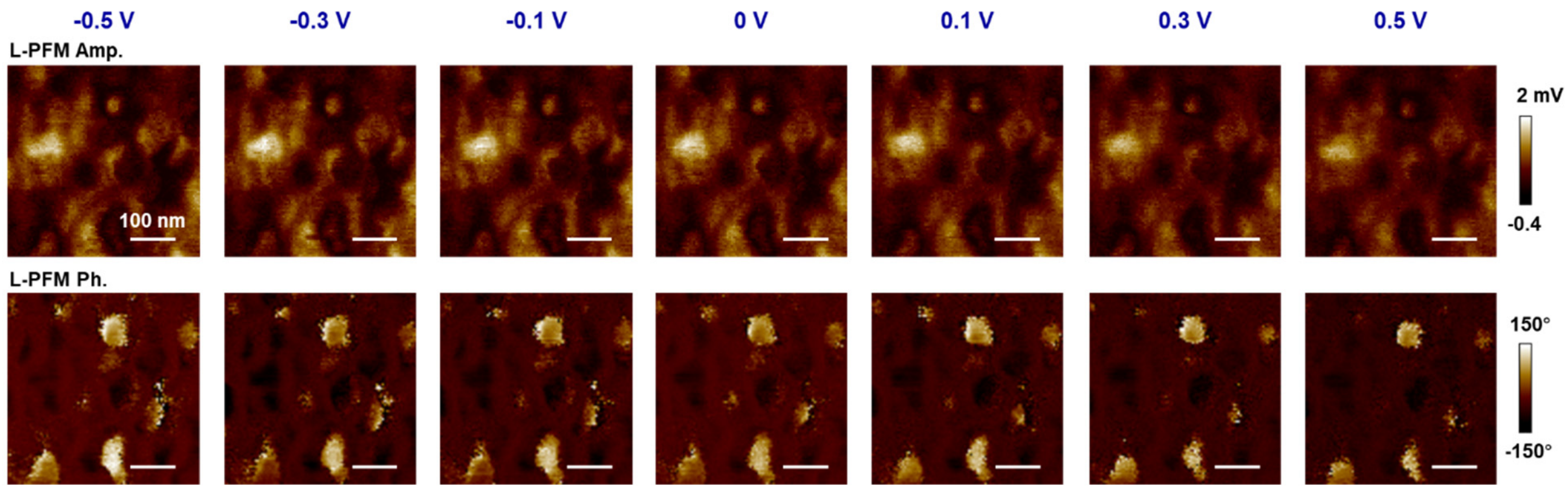

**Supplementary Figure 3 L-PFM imaging of PZT film at low $V_{bias}$.** L-PFM amplitude (top) and phase (bottom) images of the same area in a 275 nm PZT film taken at different DC bias voltages. The AC drive voltage is 100 mV. All scale bars are 100 nm.

## Supplementary Note 3: PFM Studies of Polycrystalline PZT Films at High Bias Voltage

We have also characterized the variation of local domain structure in PZT at high $V_{bias}$. Supplementary Figure 3 shows the V-PFM images taken in the same region on PZT, with $V_{bias}$ varying from 0 V to -2 V. We focus on the negative $V_{bias}$ regime to correlate with the steep slope switching behavior of the NC-FETs observed in this bias range. As shown in Supplementary Figure 4, while the domain distribution at $V_{bias}$ = -0.5 V is essentially the same as that of zero bias, the sample gradually approaches a more uniform polar state when $V_{bias}$ approaches the coercive voltage. At $V_{bias}$ = -2 V, the sample has not reached the single domain state. The evolution of domain distribution is more clearly illustrated in the L-PFM imaging (Supplementary Figure 5), which shows a pronounced reduction of DW density at $|V_{bias}| \geq 1$ V.

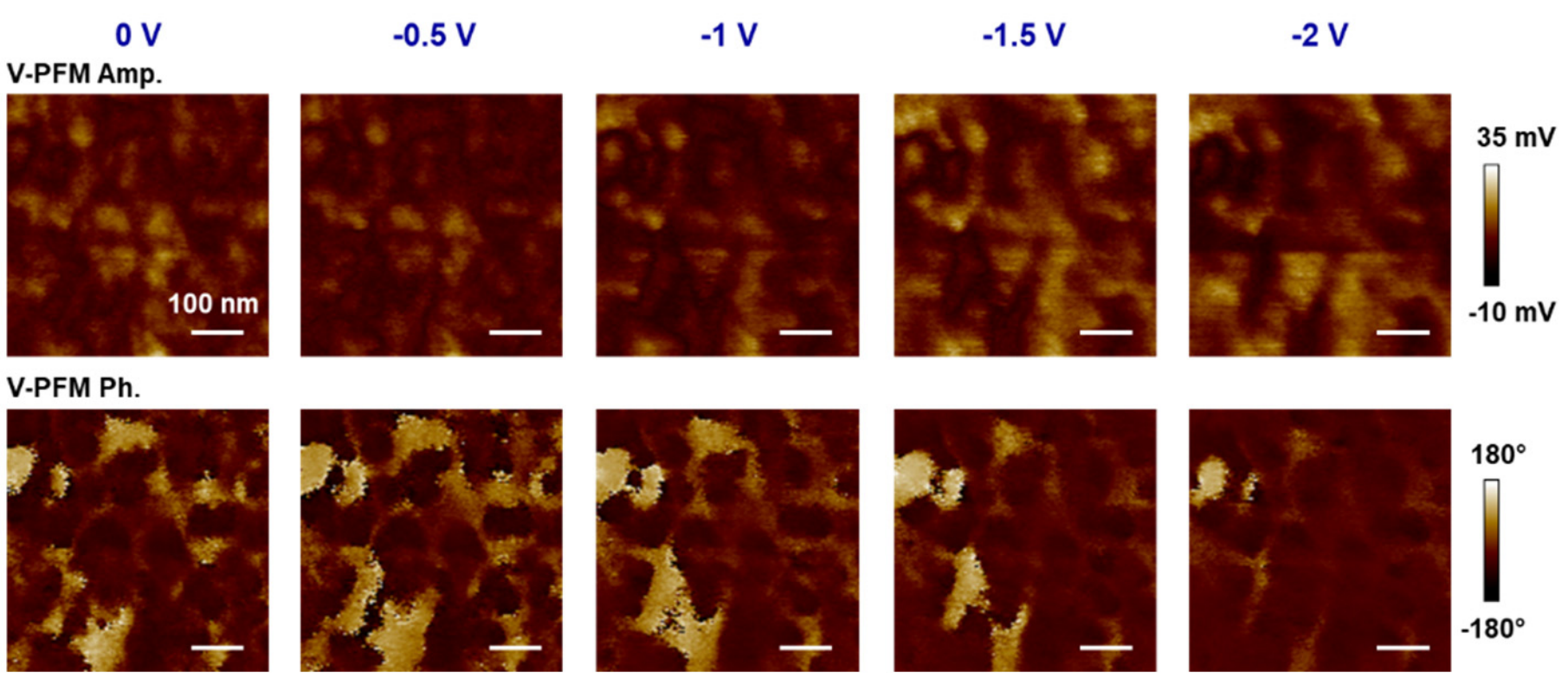

**Supplementary Figure 4 V-PFM imaging of PZT film at high $V_{bias}$.** V-PFM amplitude (top) and phase (bottom) images of the same area in a 275 nm PZT film taken at different DC bias voltages. The AC drive voltage is 100 mV. All scale bars are 100 nm.

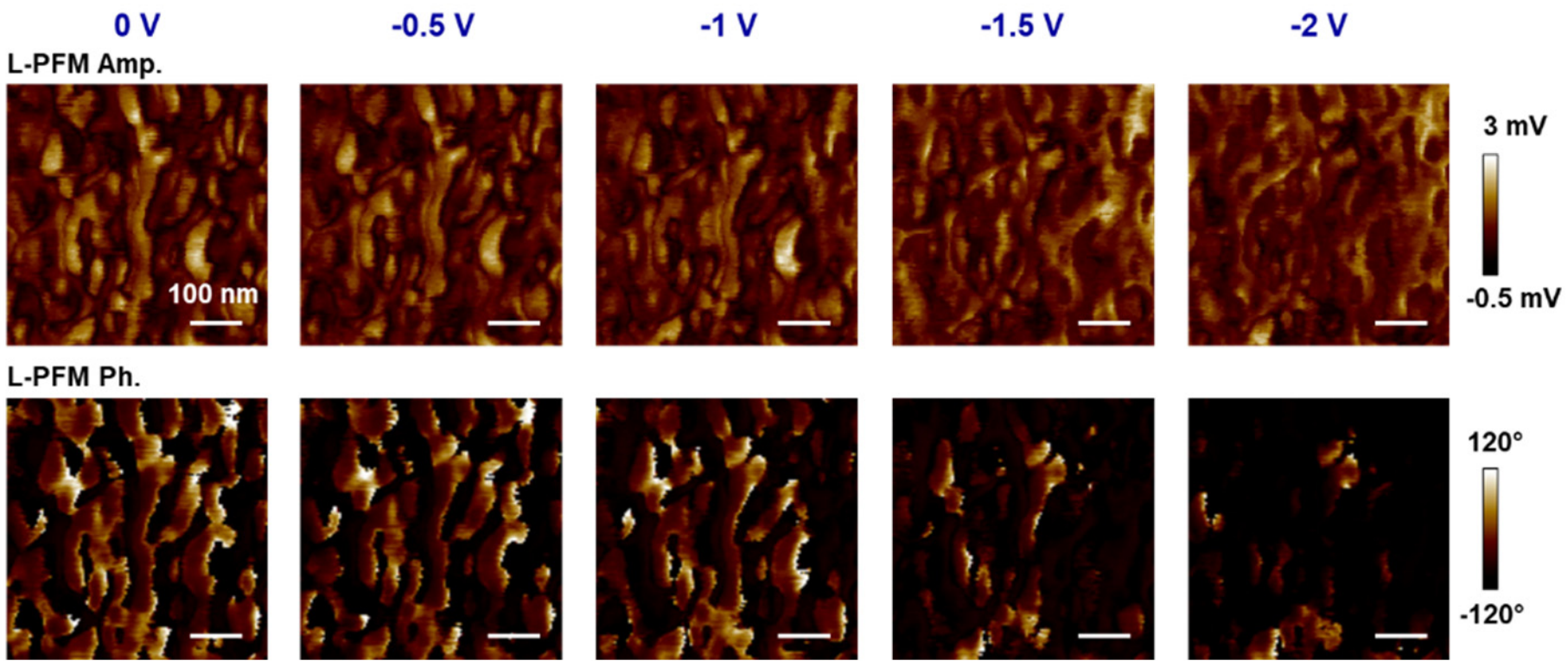

**Supplementary Figure 5 L-PFM imaging of PZT film at high $V_{bias}$.** L-PFM amplitude (top) and phase (bottom) images of the same area in a 275 nm PZT film taken at different DC bias voltages. The AC drive voltage is 100 mV. All scale bars are 100 nm.

## Supplementary Note 4: Calculation of DW Density

To gain a quantitative understanding of the DW areal density, we have mapped out the DWs in the L-PFM images. As shown in Supplementary Figure 6, we identify the DW position in the L-PFM phase image shown in Fig. 1e in the main text and extract the corresponding DW length per unit area ($l_{DW}/A$) to be 0.036±0.002 nm$^{-1}$. Previous transmission electron microscopy studies of PZT have shown that the surface chiral polar state of the DW is about 3 nm [3], based on which we estimate the DW areal density for the surface layer grains to be about 10.8%. We examine the $V_{bias}$-dependence of $l_{DW}/A$ based on the L-PFM phase images shown in Supplementary Figure 5. As shown in Fig. 5 in the main text, the DW areal density decreases from 11.8% at $V_{bias}$ = 0 V to 6.9% at $V_{bias}$ = -2 V. Note this is an under-estimate of DW density, as we only consider L-PFM images taken in one scan direction.

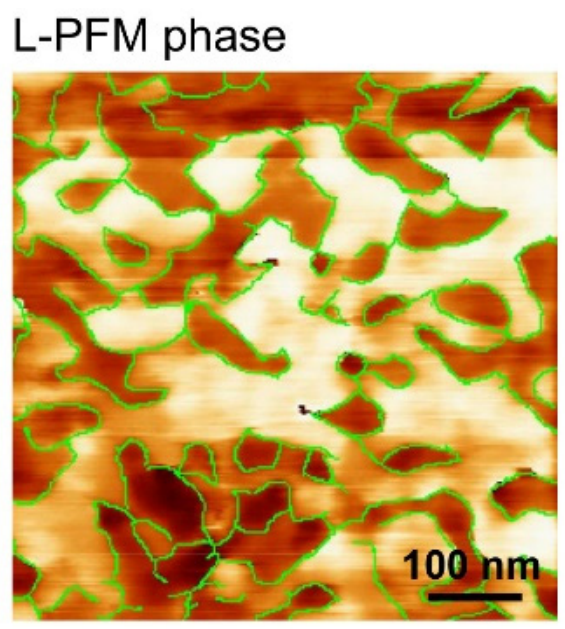

**Supplementary Figure 6 Identification of DW.** L-PFM phase image shown in Fig. 1e in the main text, with the DW positions marked by the green lines.

## Supplementary Note 5: C-AFM Studies of Leakage Current in Polycrystalline PZT Films

To track the leakage current distribution, we have carried out conductive AFM (c-AFM) studies on a thinner PZT film (150 nm) at different $V_{\text{bias}}$. As shown in Supplementary Figure 7, the film shows a uniform distribution of leakage current (Supplementary Figure 7d), which has no direct correlation with the surface morphology (Supplementary Figure 7a) and domain distributions (Supplementary Figure 7b, c). The current averaged over this area $I_{\text{ave}}$ vs. $V_{\text{bias}}$ curve exhibits a diode like behavior, similar to previous studies of DW conduction in epitaxial PZT[4], confirming the uniform leakage current distribution is real rather than due to noise.

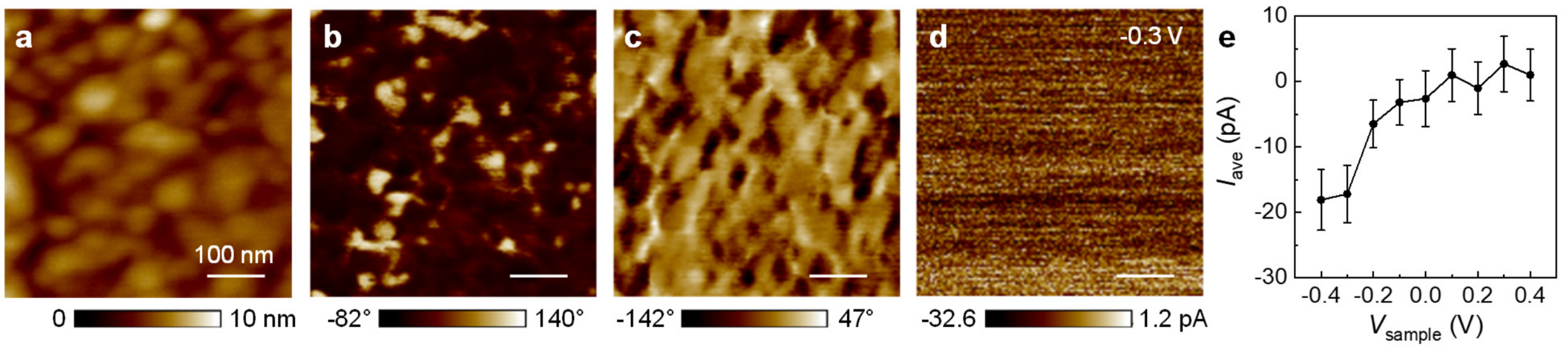

**Supplementary Figure 7 Leakage current distribution in PZT. a** AFM topography, **b** V-PFM phase image, **c** L-PFM phase image, and **d** c-AFM mapping at $V_{\text{bias}}$ = -0.3 V of the same region on a 150 nm PZT film. **e** Averaged c-AFM signal of this region as a function of $V_{\text{bias}}$. Similar behavior has been observed in multiple locations.

## Supplementary Note 6: Pyroelectric Correction to the Transfer Curves of $MoS_2$ FET at Various Temperatures

As PZT is also pyroelectric, its polarization value increases with decreasing temperature, which leads to a shift of doping level in the $MoS_2$ channel. To correct this effect in the transfer curves of the $MoS_2$ device (Fig. 2d in the main text), we have measured PZT's pyroelectric coefficient using a Au/PZT/Pt capacitor structure. As shown in Supplementary Figure 8, we slowly modulate the sample temperature at a rate of 0.1 K s$^{-1}$ between 290 K and 310 K and measure the corresponding pyroelectric current $i_{\text{pyro}}$ [5,6]. The corresponding piezoelectric coefficient is calculated using $p = \frac{i_{\text{pyro}}}{A}\left(\frac{dT}{dt}\right)^{-1}$, where $A$ is the electrode area and $t$ is time. We deduce a pyroelectric coefficient of about 60 μC m$^{-2}$ K$^{-1}$, which is smaller than previous reports of 130-200 μC m$^{-2}$ K$^{-1}$ obtained on poled polycrystalline PZT[7]. The smaller value is due to the fact that we work with unpoled PZT films within the voltage range well below the coercive voltage[8]. The doping shift can be converted to the change of gate voltage $\Delta V_g$ based on $\Delta V_g = \frac{p\Delta T d}{\varepsilon_0 \varepsilon_r}$, where $\varepsilon_r = 650$ is the dielectric constant of PZT, $\Delta T$ is the change of temperature, and $d = 300$ nm is the thickness of PZT[9,10]. We estimate a small $\Delta V_g$ of -0.03 V for lowering every 10 K in temperature. In the transfer curves shown in Fig. 2d in the main text, $V_g$ has been shifted accordingly with respect to the 300 K data. Because the subthreshold swing (SS) is calculated from the slope of the transfer curve in the quasi-linear regime, this shift of the transfer curve does not affect the SS value of the $MoS_2$ FET device.

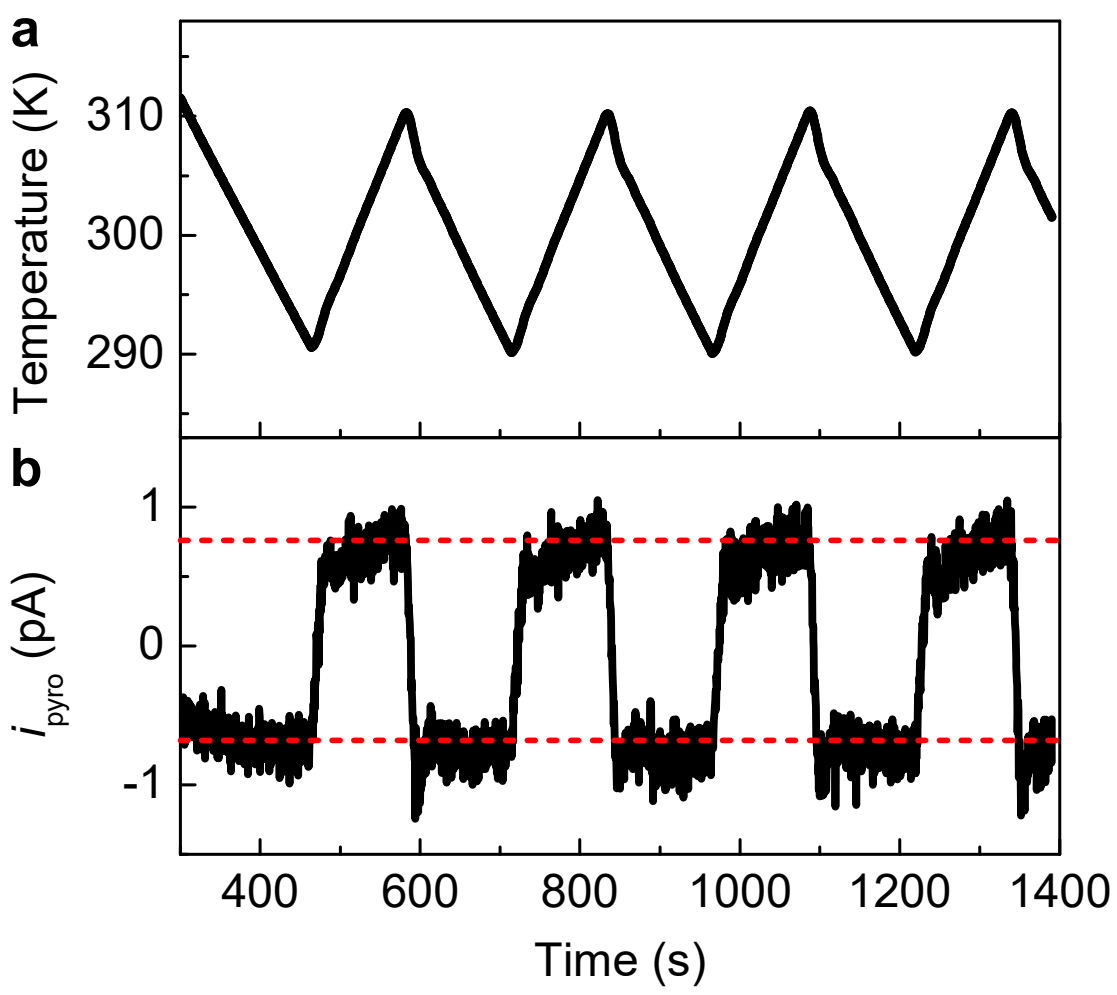

**Supplementary Figure 8  Pyroelectric coefficient measurement of PZT film. a** Temperature vs. Time profiles in the heating and cooling cycles applied to the PZT thin film, and **b** the measured pyroelectric current through PZT.

## Supplementary Note 7: Effect of Leakage Current and Extraction of Subthreshold Swing

We consider the effect of leakage current ($I_{leak}$) on the transfer characteristics of $MoS_2$ NC-FETs. In our device geometry, the source/drain contact area (~10,000 μm$^2$) is over two orders of magnitude larger than the device area (<100 μm$^2$). The leakage current is thus limited by the contacts, and the source and drain contacts are subject to different bias voltages (Supplementary Figure 9a, b). To have a controlled study of the effect of leakage current, we have performed high precision characterization of $I_{leak}$ through PZT using a capacitor structure, with the top electrode area comparable with the device contact. In Supplementary Figure 9c, we superimpose the measured $I_{leak}$ onto the transfer curve of Device 2L D2 (Fig. 3a in the main text). In the NC region, the $I_d$ curve is above the leakage level, confirming that the gate-leakage does not contribute to the steep-slope switching of the $MoS_2$ NC-FETs.

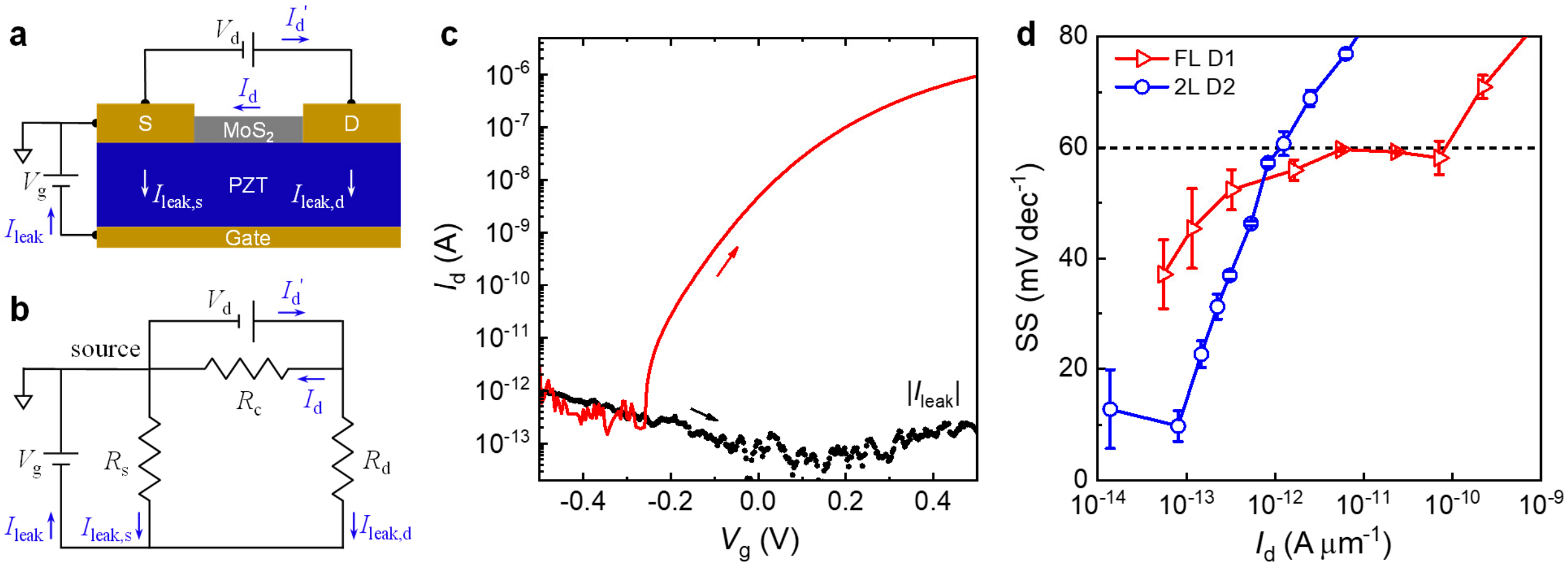

**Supplementary Figure 9 Effect of $I_{leak}$ and extraction of SS. a** Device schematic of the PZT-gated $MoS_2$ FET. **b** Equivalent circuit model for the device showing the leakage current paths. $R_c$ is the channel resistance. $R_s$ and $R_d$ are the equivalent resistance of PZT under the source and drain electrodes, respectively. $I_d$ and $I_d'$ are the actual drain current going through the $MoS_2$ channel and the measured drain current, respectively. **c** $|I_{leak}|$ vs. $V_g$ superimposed on the $I_d - V_g$ curve taken on the 2L $MoS_2$ device shown in Fig. 3a in the main text. $I_d$ here is not scaled by the channel width. **d** Averaged SS vs. $I_d$ extracted from the transfer curves for the few-layer (Fig. 2c) and 2L $MoS_2$ (Fig. 3a) devices shown in the main text. All curves are taken in the forward-scan.

The SS vs. $I_d$ data shown in Fig. 2 and Fig. 3 in the main text are obtained using the point-by-point analysis by differentiating $I_d$ at each $V_g$ point on the $I_d$-$V_g$ curve. In Supplementary Figure 9d, we show the averaged SS of the few-layer (Fig. 2c) and 2L (Fig. 3a) devices by taking the slope of any three consecutive points on the $I_d$-$V_g$ curve. The SS values are below the thermal limit for about three (two) orders of magnitude of $I_d$ for the few-layer (bilayer) device within the error bar.

## Supplementary Note 8: Analysis of the Transfer Characteristics of $MoS_2$ NC-FETs

Supplementary Figure 10a plots the transfer curves taken on a bilayer $MoS_2$ device (Device 2L D1) within $V_g = \pm 0.5$ V. With increasing $V_d$, the transfer curve shifts to positive $V_g$ direction at low current and then exhibits a crossover behavior at high current. This result is consistent with previous reports in Refs. [11,12], which can be attributed to the negative drain-induced-barrier-lowering (N-DIBL) effect in a junctionless transistor[13].

What is interesting is the direction of the shift also depends on the $V_g$ range. When the $V_g$ range is expanded to ±1 V, the transfer curve shifts to the negative $V_g$ direction for the entire gate range with no crossover behavior (Supplementary Figure 10b). Similar negative $V_g$-shift has also been reported in literature[11], which is better described as the drain-induced-barrier-lowering (DIBL) effect. This can be attributed to the junctions forming at the source/drain contacts with Ti/Au electrodes. It has been widely observed that $MoS_2$ and Ti/Au can form Schottky barrier of 50-100 meV [14,15]. The characteristic junction behavior may only manifest when the Fermi level $E_F$ of the $MoS_2$ channel is deep in the band gap. This scenario can naturally explain why DIBL is only observed at large $V_g$-range, as it corresponds to a larger polarization doping that can push $E_F$ further towards the mid-gap region. For both $V_g$-sweep ranges, the SS vs. $I_d$ curve remains independent of $V_d$ (Supplementary Figure 10c and Fig. 3b in the main text).

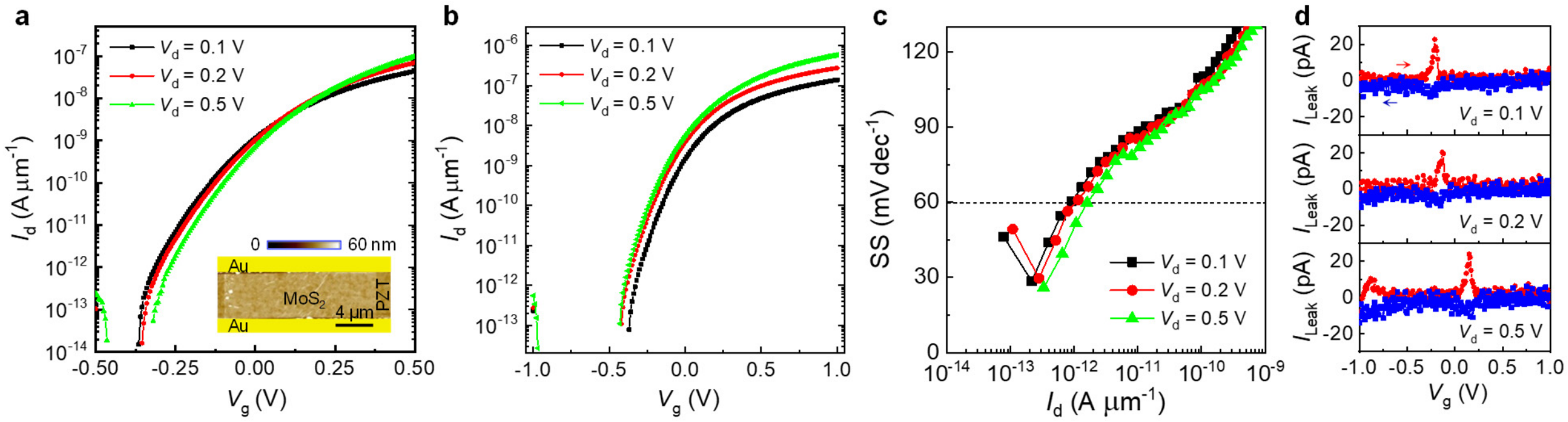

**Supplementary Figure 10 Effect of $V_d$ on the transfer characteristic for Device 2L D1. a** $I_d$ vs. $V_g$ in forward $V_g$-sweep taken at different $V_d$ values. **Inset:** AFM image of the device. The corresponding SS vs. $I_d$ is shown in Fig. 3b in the main text. **b** $I_d$ vs. $V_g$ in forward $V_g$-sweep taken at larger $V_g$-range, and **c** the corresponding SS vs. $I_d$. **d** $I_{leak}$ vs. $V_g$ taken on this device at different $V_d$. The scan rate is 10 mV s$^{-1}$.

The large $V_g$-range further allows us to examine the effect of leakage current on the transfer characteristics. Combining $V_g$ (±1 V) with $V_d$ (0.1 V), the bias applied to certain area of PZT underneath the device can approach the coercive voltage (-1.1 V). We observe clear, up to 20 pA current spikes in $I_{leak}$ (Supplementary Figure 10d), which may be due to the switching current for crystalline grains with relatively low coercive field. The number and position of the switching current peak change with $V_d$, consistent with the fact that $V_d$ causes a variation in $V_{bias}$ applied to the PZT layer underneath the $MoS_2$ channel and the source/drain electrodes (Supplementary Figure 9a, b). Despite the shift of the current spikes, the transfer curves (Supplementary Figure 10b) and corresponding SS for the forward $V_g$-sweep (Supplementary Figure 10c) remain qualitatively similar. We thus conclude that the steep slope switching observed in the forward $V_g$-sweep is not contaminated by $I_{leak}$. This current peak feature does not show up in $I_{leak}$ when the $V_g$ scan range is within $\pm 0.5\ V$, further supporting that partial polarization switching does not occur at this bias level, which is consistent with the PFM studies (Supplementary Figs. 2-3).

We have also examined the effect of $V_g$ sweep rate on the transfer characteristics. Supplementary Figure 11a shows the $I_d$-$V_g$ curves of the bilayer device (Device 2L D1) taken with scan rates of 5, 10, and 20 mV s$^{-1}$. For the hysteresis-free region ($I_d > 10^{-12}$ A μm$^{-1}$), the transfer characteristic is essentially independent of scan rate. For the forward scan, the data taken at 5 and 10 mV s$^{-1}$ exhibit consistent steep slope switching behavior. We thus conclude that the steep slope switching is an intrinsic behavior independent of scan rate. At 20 mV s$^{-1}$, the data requisition speed cannot keep up with the fast scan rate, with the problem amplified by the rapid variation of $I_d$ in the steep slope switching regime. As a result, there is a forward shift of the $I_d$ vs. $V_g$ curve, and we have only collected a limited number of data points in the sub-60 mV dec$^{-1}$ region.

For the reverse scan, there is a clear change of the saturation current level, with the current floor increasing quasi-linearly with the scan rate. It suggests that the current measured in this regime is likely due to extrinsic mechanisms. Possible culprits include the charge dynamics of interfacial adsorbates (*e.g.*, water dissociation and recombination) and charge trapping/de-trapping. It has been shown in previous studies that such interfacial charge dynamics can compromise the retention and switching in van der Waals FETs back-gated by ferroelectric oxides, and can even induce anti-hysteresis[9,16,17]. Based on the sweeping speed dependence, we believe that the hysteresis at low current level can be attributed to these extrinsic mechanisms, while the sub-60 mV dec$^{-1}$ switching in the forward scan is intrinsic to the ferroelectric system.

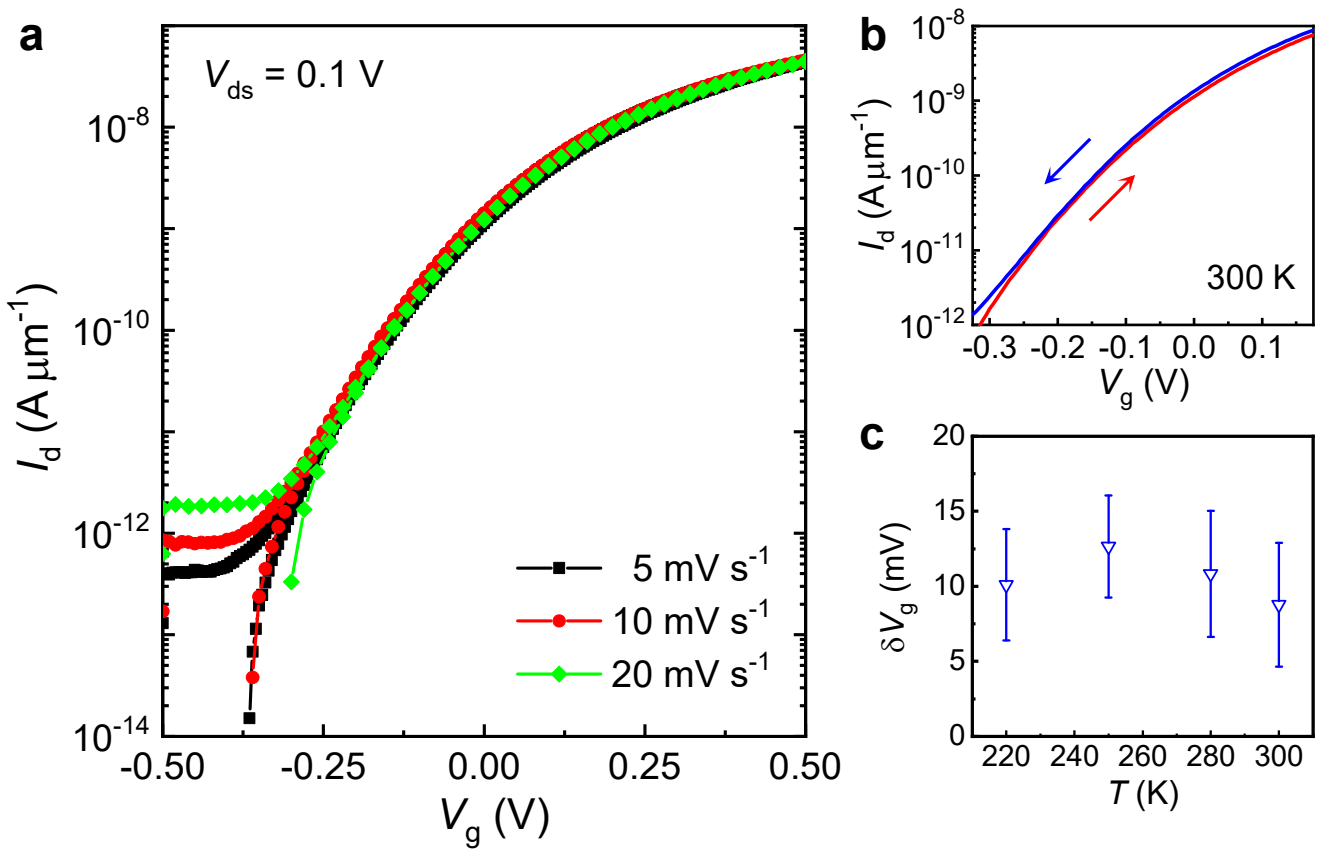

**Supplementary Figure 11 Effects of $V_g$ scan rate and temperature on the transfer characteristic.** **a** $I_d$ vs. $V_g$ for Device 2L D1 taken at different $V_g$ scan rates. The corresponding SS vs. $I_d$ is shown in Fig. 3c in the main text. **b** Expanded view of the 5 mV s$^{-1}$ data to illustrate the extraction of $\delta V_g$. **c** Averaged $V_g$ hysteresis window vs. temperature.

We have also examined the temperature-dependence of the switching hysteresis. As shown in Supplementary Figure 11b, we extract the $V_g$ hysteresis window ($\delta V_g$) between the forward-sweep and reverse-sweep transfer curves of Device 2L D1. Supplementary Figure 11c shows $\delta V_g$ averaged over 4 orders of magnitude of $I_d$ ($10^{-12}$-$10^{-8}$ A $\mu m^{-1}$) at four different temperatures (220 to 300 K) below the Curie temperature of PZT (above 700 K), illustrating that $\delta V_g$ is nearly temperature independent. This result indicates that the steep slope switching of the $MoS_2$ FET is not due to the net effect of competing mechanisms, *e.g.*, ferroelectric switching and interfacial charge dynamics, as those would have differen temperature dependences. It further supports our conclusion that the NC effect is intrinsic to the polycrystalline PZT gate.

**Supplementary Note 9: High Field Hysteresis in the Transfer Characteristics of $MoS_2$ NC-FETs**

When scanning $V_g$ in a large voltage-range (*e.g.*, $|V_g| > 1$ V), a switching hysteresis occurs in the $MoS_2$ NC-FETs. Supplementary Figure 12a, b shows the transfer characteristics of a bilayer $MoS_2$ NC-FET (Device 2L D3) taken at the $V_g$-scan ranges of $\pm$1.5 V and $\pm$2 V, respectively. The device exhibits clockwise switching hysteresis similar to those reported in literature[18]. The hysteresis window increases with increasing $V_g$-range, suggesting that it is dominated by the extrinsic mechanism[9]. A clockwise hysteresis has been widely observed in $MoS_2$ FETs back-gated by dielectrics, which can be attributed to the dynamic response of ambient adsorbates, such as water dissociation/recombination, or charge trapping/de-trapping in interfacial defect states[9,19]. For ferroelectric-gated $MoS_2$ FET, both clockwise and anti-clockwise hysteresis have been observed due to the competition between the intrinsic and extrinsic mechanisms. Previous studies have shown that for $MoS_2$ devices prepared on PZT with relatively rough surfaces, the hysteresis is clockwise[18], which is consistent with the results in this study.

The anti-hysteresis, on the other hand, is also absent in the low $V_g$-range, as the dynamic response of interfacial charges can only be activated above a threshold voltage [19,20]. A control study has been performed on a bilayer $MoS_2$ FET back-gated by 300 nm $SiO_2$ (Device 2L D4). As shown in Supplementary Figure 12c, clockwise hysteresis in the $I_d$-$V_g$ curve is observed in this device at high $V_g$-scan range ($\pm$3 V and above), while there is no hysteresis at low $V_g$-scan range ($\pm$1 V), confirming this scenario. We also note that no steep slope switching is observed in this device ($SS_{min}$ = 106 mV $dec^{-1}$).

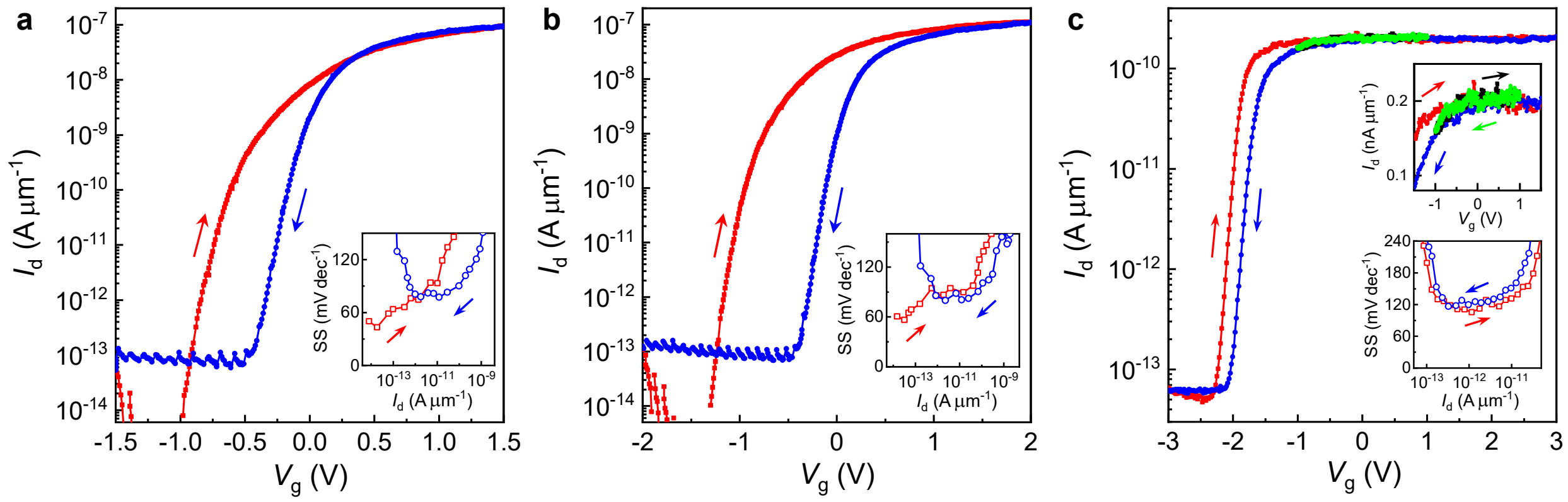

**Supplementary Figure 12 Switching hysteresis in the transfer characteristic at high $V_g$-range. a, b** Transfer characteristics of a bilayer $MoS_2$ FET (Device 2L D3) at **a** $V_g$-range of ±1.5 V, and **b** $V_g$-range of ±2 V. **Insets:** The corresponding SS vs. $I_d$. **c** Transfer characteristics of a 2L $MoS_2$ FET (Device 2L D4) back-gated by 300 nm $SiO_2$ taken at the $V_g$-ranges of ±1 V (green and black symbols) and ±3 V (blue and red symbols) and $V_d$ = 0.1 V. Insets: (Top) Expanded view of the transfer curves; (Bottom) the corresponding SS vs. $I_d$.

We can thus conclude that the hysteresis-free operation at $I_\mathrm{d} > 10^{-12}$ A µm$^{-1}$ observed in the $MoS_2$ NC-FET is intrinsic to the polycrystalline PZT film at low $V_g$ below the coercive voltage. Despite the hysteresis at high $V_g$-range, sub-60 mV dec$^{-1}$ SS is also observed in the forward $V_\mathrm{g}$-sweep (Supplementary Figure 12a, b inserts), with $SS_{min}$ increasing with increasing $V_g$-scan range (Fig. 5b in the main text). This is consistent with the PFM studies, which show that the PZT film has not reached the single domain state at $V_{bias}$ = -2 V (Supplementary Figs. 4-5).

## Supplementary Note 10: 3D Force Field Simulation

The 3D force field simulation is mainly based on the model developed in Refs. [21,22]. The single crystal term $f_\mathrm{bulk}$ in the Gibbs free energy $F$ of the multi-domain structure (Equation (2) in the main text) can be expressed as a 6$^{th}$-order polynomial with respect to the polarization along three Cartesian-axes as[21,22]:

$$f_\mathrm{bulk} = \alpha_1(P_1^2 + P_2^2 + P_3^2) + \alpha_{11}(P_1^4 + P_2^4 + P_3^4) + \alpha_{12}(P_1^2P_2^2 + P_2^2P_3^2 + P_3^2P_1^2) + \alpha_{123}P_1^2P_2^2P_3^2 + \alpha_{111}(P_1^6 + P_2^6 + P_3^6) + \alpha_{112}[P_1^4(P_2^2 + P_3^2) + P_2^4(P_3^2 + P_1^2) + P_3^4(P_1^2 + P_2^2)] \, , \tag{1}$$

The gradient term $f_\mathrm{grad}$ captures the energy contribution from the polarization discontinuity in domain walls and surfaces:

$$f_{\text{grad}} = \frac{1}{2}G_{11}\left(P_{1,1}^2 + P_{2,2}^2 + P_{3,3}^2\right) + G_{12}\left(P_{1,1}P_{2,2} + P_{2,2}P_{3,3} + P_{3,3}P_{1,1}\right) + G_{44}\left(P_{1,2}^2 + P_{2,1}^2 + P_{2,3}^2 + P_{3,2}^2 + P_{1,3}^2 + +P_{3,1}^2\right) + f_{\text{surf}} , \quad (2)$$

where $P_{i,j} = \partial P_i/\partial x_j$, and $G_{i,j}$ are the gradient coefficients. The $f_{\text{surf}}$ term describes the energy cost due to the discontinuity of electric displacement at the surfaces of the ferroelectric insulator:

$$f_{\text{surf}} = G_{\text{S}}(P_3 - Q)^2|_{x_3=0,\, d_3} , \quad (3)$$

where $Q$ is the interface charge density and $d_3$ is the thickness of the ferroelectric insulator. Therefore, $x_3 = d_3$ is the position of channel/insulator interface, and $x_3 = 0$ is the insulator/substrate interface. The elastic term

$$f_{\text{elas}} = \frac{1}{2}C_{11}(e_{11}^2 + e_{22}^2 + e_{33}^2) + C_{12}(e_{11}e_{22} + e_{22}e_{33} + e_{33}e_{11}) + 2C_{44}(e_{12}^2 + e_{23}^2 + e_{31}^2) , \quad (4)$$

where $C_{ij}$ are the elastic constants in the Voigt's notation, and

$$e_{ij} = \varepsilon_{ij} - \varepsilon_{ij}^0 . \quad (5)$$

Here $\varepsilon_{ij}$ are strains, and

$$\begin{aligned}
\varepsilon_{11}^0 &= Q_{11}P_1^2 + Q_{12}(P_2^2 + P_3^2) , \\
\varepsilon_{22}^0 &= Q_{11}P_2^2 + Q_{12}(P_3^2 + P_1^2) , \\
\varepsilon_{33}^0 &= Q_{11}P_3^2 + Q_{12}(P_1^2 + P_2^2) , \\
\varepsilon_{12}^0 &= Q_{44}P_1P_2 , \\
\varepsilon_{23}^0 &= Q_{44}P_2P_3 , \\
\varepsilon_{13}^0 &= Q_{44}P_1P_3
\end{aligned} \quad (6)$$

are the stress-free strain for the structure with a certain polarization, with $Q_{ij}$ the electrostrictive coefficients. Since the PZT film is grown on a Pt electrode, we consider it as a clamped multi-crystalline film with zero strain.

The electrostatic term:

$$f_{\text{elec}} = -E_1P_1 - E_2P_2 - E_3P_3 . \quad (7)$$

Along the in-plane direction, we use the periodic boundary and short circuit conditions[22], in which the following electrostatic restriction should be satisfied:

$$2V_{\text{e1}} + V_{\text{o1}} = 0 ,$$

$$2V_{\mathrm{e}2} + V_{\mathrm{o}2} = 0\ , \tag{8}$$

$$V_{\mathrm{e}3} + V_{\mathrm{o}3} + \psi_{\mathrm{s}} = V_{\mathrm{g}}\ ,$$

where $V_{\mathrm{e1,e2,e3}}$ and $V_{\mathrm{o1,o2,o3}}$ are the potential drop through the electrode and ferroelectric oxide, respectively, along the *x*, *y* and *z* directions. Our previous work has shown that:

$$V_{\mathrm{e}i} = \frac{Q_i \lambda_i}{\epsilon_{\mathrm{e}} \epsilon_0}\ , \tag{9}$$

$$V_{\mathrm{o}i} = E_i \cdot d_i = \frac{Q_i - P_i}{\epsilon_0} d_i\ ,$$

where $Q_i$ is the charge density in the electrodes or at the surfaces, $\lambda_{\mathrm{i}}$ is the screening length, and $d_i$ is the thickness of the insulators along the *i* (= 1, 2, 3) direction[22]. Therefore, Supplementary Equation (7) can be rewritten as:

$$\frac{2Q_1\lambda_1}{\epsilon_{\mathrm{e}}\epsilon_0} + \frac{Q_1 - P_1}{\epsilon_0} d_1 = 0\ ,$$

$$\frac{2Q_2\lambda_2}{\epsilon_{\mathrm{e}}\epsilon_0} + \frac{Q_2 - P_2}{\epsilon_0} d_2 = 0\ , \tag{10}$$

$$\frac{Q_3\lambda_3}{\epsilon_{\mathrm{e}}\epsilon_0} + \frac{Q_3 - P_3}{\epsilon_0} d_3 + \psi_{\mathrm{s}}(Q_3) = V_{\mathrm{g}}\ .$$

Deduced from Supplementary Equation (10), we have:

$$E_1 = -\frac{2\lambda_1 P_1}{(2\lambda_1 + d_1\epsilon_{\mathrm{e}})\epsilon_0}\ ,$$

$$E_2 = -\frac{2\lambda_2 P_2}{(2\lambda_2 + d_2\epsilon_{\mathrm{e}})\epsilon_0}\ ,$$

$$E_3 = -\frac{Q_3 - P_3}{\epsilon_0}\ ,$$

$$E_3 = -\frac{1}{d_3}\left[V_{\mathrm{g}} - \frac{Q_3\lambda_3}{\epsilon_{\mathrm{e}}\epsilon_0} - \psi_{\mathrm{s}}(Q_3)\right]\ . \tag{11}$$

The relationship between $Q_3$ and $\psi_s$ is

$$Q_3 = \sqrt{2\epsilon_{\mathrm{MoS_2}} k_{\mathrm{B}} T N_{\mathrm{a}}} \cdot \left[\left(e^{\frac{-q\psi_{\mathrm{s}}}{k_{\mathrm{B}}T}} + \frac{q\psi_{\mathrm{s}}}{k_{\mathrm{B}}T} - 1\right) + \frac{n_i^2}{N_{\mathrm{a}}^2}\left(e^{\frac{q\psi_{\mathrm{s}}}{k_{\mathrm{B}}T}} - \frac{q\psi_{\mathrm{s}}}{k_{\mathrm{B}}T} - 1\right)\right]^{\frac{1}{2}}\ , \tag{12}$$

where $\epsilon_{\mathrm{MoS_2}}$ is the dielectric constant, $n_i$ is the intrinsic carrier concentration, and $N_{\mathrm{a}}$ is the doping concentration[23]. In the simulation, we use a 20×20×10 supercell. The polarization of each unit cell is initially randomly directed. Then, the Gibbs free energy *F* is minimized by the Landau-Khalatnikov equation:

$$\gamma \frac{d\vec{P}}{dt} + \nabla_{\vec{P}} F = 0\ , \tag{13}$$

where $\gamma$ is the polarization dynamic parameter[24]. After acquiring the equilibrated structure, $V_g$ switches as 0 → 0.5 → 0 → -0.5 → 0 V with a 0.05 V step. For each voltage, the structure acquired at the last step is used as the beginning structure, and then $F$ is minimized by Supplementary Equation (13).

The drain-source current $I_d$ is obtained by the Pao-Sah double integral as:

$$I_{\mathrm{d}} = q\mu_{\mathrm{eff}}\frac{W}{L}\int_0^{V_{\mathrm{d}}}\left[\int_{\delta}^{\psi_{\mathrm{s}}-V_{\mathrm{FB}}}\frac{n_i^2}{N_{\mathrm{a}}}\frac{e^{q(\psi-V)/k_{\mathrm{B}}T}}{E(\psi,V)}d\psi\right]dV + I_0\ , \tag{14}$$

where $q$ is the electronic charge, $\mu_{\mathrm{eff}}$ is the effective electron mobility, $W$ is the width of the channel, $L$ is the length of the channel, $V_{\mathrm{FB}}$ is the flat band potential, $I_0$ is the current floor imposed by extrinsic mechanisms, such as interfacial charge dynamics or measurement noise level, $E(\psi, V)$ is the electric field in the channel, and $\delta$ is an infinitesimal quantity[25]. The parameters involved in this study are listed in Supplementary Table 2.

**Supplementary Table 2 Parameters involved in the 3D force field simulation.**

| | Description | Value |
|---|---|---|
| $T$ | Temperature | 300 K |
| $\alpha_1$ | Coefficient in LGD theory[a] | $-1.06\times10^{8}$ m F$^{-1}$ |
| $\alpha_{11}$ | Coefficient in LGD theory[a] | $2.39\times10^{7}$ m$^{5}$ C$^{-2}$ F$^{-1}$ |
| $\alpha_{12}$ | Coefficient in LGD theory[a] | $4.25\times10^{8}$ m$^{5}$ C$^{-2}$ F$^{-1}$ |
| $\alpha_{111}$ | Coefficient in LGD theory[a] | $2.17\times10^{8}$ m$^{9}$ C$^{-4}$ F$^{-1}$ |
| $\alpha_{112}$ | Coefficient in LGD theory[a] | $9.75\times10^{9}$ C$^{-4}$ F$^{-1}$ |
| $\alpha_{123}$ | Coefficient in LGD theory[a] | $-4.70\times10^{9}$ m$^{9}$ C$^{-4}$ F$^{-1}$ |
| $G_{11}$ | Gradient coefficient[c] | $4.41\times10^{-10}$ C$^{-2}$ m$^{4}$ N |
| $G_{12}$ | Gradient coefficient[c] | 0 |
| $G_{44}$ | Gradient coefficient[c] | $2.07\times10^{-10}$ C$^{-2}$ m$^{4}$ N |
| $\Delta x_{1,2,3}$ | grid spacing in calculating the gradient term[c] | 1 nm |
| $G_{\mathrm{S}}$ | Surface gradient coefficient[c] | $G_{11}/\Delta x_3$ |
| $C_{11}$ | Elastic constants[b] | 1.746 N m$^{-2}$ |
| $C_{12}$ | Elastic constants[b] | 0.7937 N m$^{-2}$ |
| $C_{44}$ | Elastic constants[b] | 1.11 N m$^{-2}$ |
| $Q_{11}$ | Electrostrictive coefficients[a] | $7.8\times10^{-2}$ C m$^{-4}$ |
| $Q_{11}$ | Electrostrictive coefficients[a] | $-2.61\times10^{-2}$ C m$^{-4}$ |
| $Q_{11}$ | Electrostrictive coefficients[a] | $3.22\times10^{-2}$ Cm$^{-4}$ |

| $\lambda_{1,2,3}$ | Screening lengths in electrodes[a] | 0.2 nm |
|---|---|---|
| $d_{1,2,3}$ | Dimensions of the nano-region in our simulation[d] | 300 nm |
| $\epsilon_e$ | Dielectric constants of electrodes | 1.0 F m$^{-1}$ |
| $\epsilon_{MoS_2}$ | Dielectric constant of $MoS_2$ | 4.2 F m$^{-1}$ |
| $N_a$ | Substrate doping concentration | 5.0×10$^{11}$ cm$^{-3}$ |
| $n_i$ | Intrinsic carrier concentration | 1.0×10$^{3}$ cm$^{-3}$ |
| $\mu_{eff}$ | Effective electron mobility[d] | 8 cm$^2$ V$^{-1}$ s$^{-1}$ |
| $W$ | Width of the channel[d] | 7 µm |
| $L$ | Length of the channel[d] | 3.2 µm |
| $V_{FB}$ | Flat band potential | 1.3 V |
| $\gamma$ | Polarization dynamic parameter[c] | 0.01/$\alpha_1$ |

[a] Ref. [26]

[b] Ref. [22]

[c] Estimated from Ref. [22]

[d] Measured in the current work

**Supplementary Note 11: Energy for Domain Wall Motion**

We use a simple model to demonstrate that the energy and electric field required to polarize the metastable states at DWs are much smaller than those for reorienting the polarization of the entire crystal. Here, we consider a one-dimensional (1D) PZT crystal under zero strain, which is composed of 99 unit cells. The center of the DW locates between the #50 and #51 cells, as shown in Supplementary Figure 13a. The free energy of the system is expressed as:

$$F = \sum_i \left[\alpha_1 P_i^2 + \alpha_{11} P_i^4 + \alpha_{111} P_i^6 + G(P_i - P_j)^2\right], \quad (15)$$

where $i$ and $j$ are neighboring cells. Supplementary Figure 13b shows the calculated polarization profile via minimizing the free energy. Starting from the DW center, the polarizations of the first three up-polarized unit cells are marked as $P_1$, $P_2$, and $P_3$, respectively. Since the domain structure is anti-symmetric, the polarizations of the first three down-polarized unit cells are $-P_1$, $-P_2$, and $-P_3$, respectively. We consider $P_\infty = 0.5$ C m$^{-2}$ as the bulk polarization in a uniform domain, and in our simulation $P_\infty \approx P_3$. Next, we artificially change the polarization of the #50 unit cell $P(\#50)$ and optimize the polarization of the other unit cells with the Landau-Khalatnikov equation (Supplementary Equation (13)).

Supplementary Figure 13c shows the calculated energy profile as a function of $P(\#50)$. As expected, the conditions of $|P(\#50)| = \pm P_1$, $\pm P_2$, and $\pm P_3$ give the same energy (about 3 meV), since they correspond to either a left (+) or right (-) shift of the DW, with the resulting domain structures correlated by the translational symmetry. If the depolarization field is taken into consideration, the free energy is rewritten as:

$$F = \sum_i \left[\alpha_1 P_i^2 + \alpha_{11} P_i^4 + \alpha_{111} P_i^6 + G(P_i - P_j)^2\right] + \beta \bar{P}^2, \tag{16}$$

where $\beta$ is a constant depending on the screening length and thickness, and $\bar{P}$ is the average polarization. As shown in Supplementary Figure 13d, the less-polarized states $|P(\#50)| = \pm P_1$ become more favorable, and the energy corresponds to the condition $|P(\#50)| = \pm P_3$ increases to about 10 meV. In contrast, Supplementary Figure 13e shows the double well energy of a uniformly polarized single crystal. The energy barrier between the two polarization states is about 600 meV, more than two orders of magnitude higher than those for the DW motion considered in Supplementary Figure 13c, d. We note that the polarization value $\pm P_1$ is within the NC regime for polarization reversal (Supplementary Figure 13e), confirming that the metastable polar states at the DWs are stabilized in the NC mode.

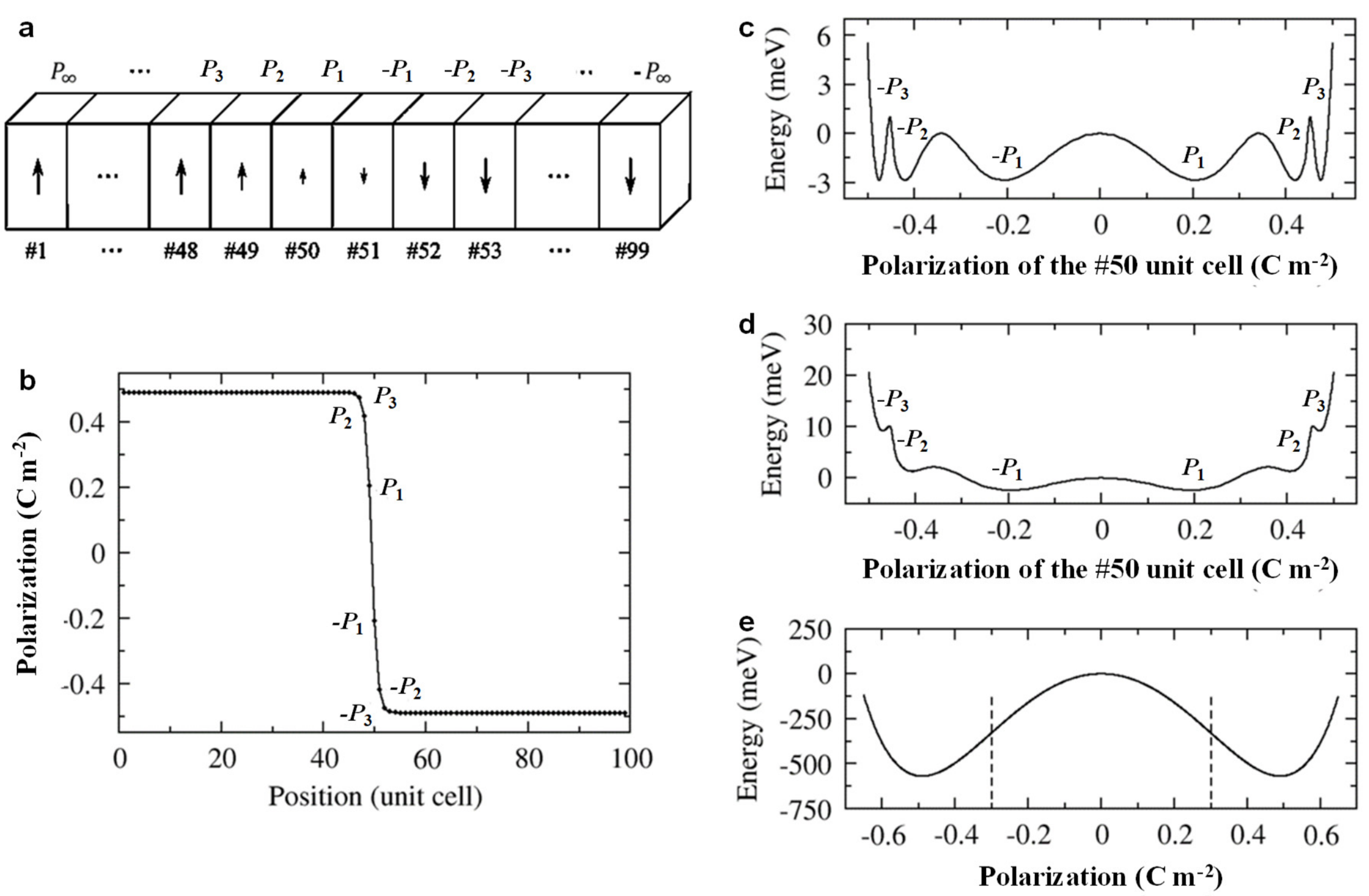

**Supplementary Figure 13 Theoretical modeling of the threshold energy required for polarization change at the DW. a** Schematic model of a 1D PZT crystal composed of 99 unit cells. It hosts a 180°

DW. **b** Polarization profile in the 1D crystal. **c, d** Energy of the 1D crystal as a function of the polarization of the #50 unit cell modeled using Supplementary Equation (15) (**c**) and Equation (16) (**d**). **e** Energy double well of the single crystal. The segment between the two dashed lines corresponds to the NC states.

The total energy barrier for DW motion has to be scaled by the width of the DW, as the entire DW has to shift collectively. Since the 1D model does not take into account the long-range dipole-dipole interaction for a two-dimensional DW, the DW motion energy in a three-dimensional (3D) ferroelectric should be much larger. On the other hand, this simplified 1D model clearly illustrates that the energy barrier for polarizing the DW region is much smaller than polarization reversal for a uniformly polarized domain, which can be generalized to the 3D ferroelectric systems.

## Supplementary References